\documentclass[preprint,showpacs,preprintnumbers]{revtex4}
\usepackage{amsmath}
\usepackage{graphicx}
\usepackage{dcolumn}
\usepackage{bm}
\usepackage{epsfig}
\usepackage[T1]{fontenc}
\usepackage{ae,aecompl}
\usepackage{epstopdf, epsfig}
\usepackage{color}
\usepackage{appendix}

\begin{document}

\title{Nonequilibrium and morphological characterizations of
	Kelvin-Helmholtz instability in compressible flows}
%\title{Effects of viscosity and heat conduction on Kelvin-Helmholtz instability:
%Discrete Boltzmann study, non-equilibrium and morphological characterizations}
\author{Yanbiao Gan$^{1,2}$, Aiguo Xu$^{3,4,5}$\footnote{
Corresponding author. E-mail: Xu\_Aiguo@iapcm.ac.cn}, Guangcai Zhang$^3$,
Chuandong Lin$^{6}$, Huilin Lai$^{2}$, Zhipeng Liu$^{7}$}
\affiliation{1, North China Institute of Aerospace Engineering, Langfang 065000, China\\
2, College of Mathematics and Informatics $\&$ FJKLMAA, Fujian Normal
University, Fuzhou 350007, China \\
3, Laboratory of Computational Physics, Institute of Applied
Physics and Computational Mathematics, P. O. Box 8009-26, Beijing 100088,
China \\
4, Center for Applied Physics and Technology, MOE Key Center for High Energy
Density Physics Simulations, College of Engineering, Peking University,
Beijing 100871, China\\
5, Center for Applied Physics and Technology, MOE Key Center for High Energy Density Physics Simulations, College of Engineering, Peking University, Beijing 100871, China\\
6, Center for Combustion Energy, Key Laboratory for Thermal Science and Power Engineering of Ministry of Education, Department of Energy and Power Engineering, Tsinghua University, Beijing 100084, China\\
7, Department of Physics, School of Science, Tianjin Chengjian University, Tianjin, 300384\\
}
\date{\today }

\begin{abstract}
We investigate the effects of viscosity and heat conduction on the onset and
growth of Kelvin-Helmholtz instability (KHI) via an efficient discrete
Boltzmann model. Technically, two effective approaches are presented to quantitatively analyze and understand the configurations and kinetic processes.
One is to determine the thickness of mixing layers through tracking the distributions and evolutions of the thermodynamic nonequilibrium (TNE) measures; the other is to evaluate the growth rate of KHI from the slopes of morphological functionals. Physically, it is found that the time histories of width of mixing layer, TNE intensity, and boundary length show high correlation and attain their maxima simultaneously. The viscosity effects are twofold, stabilize the KHI, and enhance both the local and global TNE intensities. Contrary to the monotonically inhibiting effects of viscosity, the heat conduction effects firstly refrain then enhance the evolution afterwards. The physical reasons are analyzed and presented.
\end{abstract}

\preprint{}
\maketitle

\section{Introduction}

 The Kelvin-Helmholtz instability (KHI) occurs at a perturbed interface
between two fluids or two parts
of the same fluid with different tangential velocities \cite{1}. As
an efficient and important
initiating mechanism of turbulence and mixing of fluids \cite{2,3,4,5,6,7}, it plays crucial roles in various fields, ranging from high-energy-density
physics \cite{8}, geophysics and astrophysics \cite{9,10,11,12,13,14}, inertial confinement fusion (ICF) \cite{15,16,17},
combustion \cite{18,19,20},
to Bose-Einstein condensate \cite{21,22}
and graphene \cite{23}, etc.
Concretely, in geophysical and astrophysical situations, on the one hand,
the fully developed KH billows are responsible for the formation of
large-scale vortical structures in systems such as hurricane \cite{9},
galaxy spiral arms \cite{10}, heliopause \cite%
{11,12},
and solar wind interaction with the Earth's magnetosphere \cite
{13,14}, leading to violent
intermixing across shear layers; on the other hand, the significantly
suppressed KH roll-ups
contribute to the sufficiently long, stable
and highly collimated supersonic astrophysical jets \cite%
{16,24,25,26} with
length-to-width ratios as high as $100$ or more, emanated from young stellar
objects or active galactic nuclei \cite{27},
and jet-like long spikes observed in the high-energy-density laboratory
astrophysics experiments \cite{28}.
In ICF, at the final stage of the Rayleigh-Taylor instability (RTI) \cite
{29,30,31,32}, KHI is triggered at the spike tips due to
the relative motion of light and heavy components. As a secondary
instability, the appearance of KHI aggravates the development of
nonlinearity of RTI, quickens the mixing of fluid on small-scale, and
produces mushroom shaped structures around the interface \cite
{15,16,33,34}.
In combustion, intense KHI induced by the interaction between explosion and
flame at the late stage of deflagration-to-detonation transition has been
revealed analytically and experimentally \cite{19}. The
resulted KHI makes an accelerated turbulent burning that enhances the
reaction process of the fuel mixture and leads to the detonation triggering
finally.

Owing to its extreme importance in fields including, but not limited to,
listed above,
the KHI has been investigated extensively
through experiments, theoretical analysis and more recently by numerical
simulations during the past decades \cite{35,36,37,38,39,40,41,42,43,44,45,46,47,48,49,50,51,52,53,54}. Those studies indicate the following scenarios. During its life-cycle,
a small perturbation whose wavelength along the interface between two fluids
is longer than the shear layer width, will experience
linear and nonlinear growth stages to a saturation point followed by a
post-saturation evolution, whereafter, may evolve into turbulent mixing
through mass-momentum-energy transport and cascades of interacting vortices.
Those studies also demonstrate that density ratio \cite%
{35,36,37,38,39}, viscosity \cite{39,40,41}, surface tension
\cite{38,39,42}, magnetic field \cite%
{43,44}, and compressibility \cite{37,43,45,46} suppress the
evolution, while the velocity difference \cite
{38,47}, and the density transition layer \cite
{47,48} favor the evolution. Despite these significant
progresses to date,
there still remains several fundamental issues that deserve special attention.

\emph{First, the kinetic modeling of KHI needs further investigation, and the thermodynamic nonequilibrium (TNE) effects needs careful consideration.} It has been stressed that kinetic effects will come into play and the traditional fluid modeling is not sufficient for complex fluids when the characteristic scale becomes small, so that the Knudsen number becomes large, and to study the thermodynamic nonequilibrium (TNE) effects is one of the key means to investigate the fundamental kinetic processes
\cite{55,56,57,58,59,60,61,62,63,64,65,66,67,68,69,70,71,72,73,74,75}.
Examples for KHI in plasma are referred to Refs. \cite{76,77}.
While, in previous studies, nearly all numerical investigations were based on hydrodynamic equations at the Euler level \cite{8,10,18,25,26,33,34,35,36,37,48}
which assume that the system is always in its local thermodynamic equilibrium.

We further stress that the TNE manifestations are significant, even dominant during the development of KHI, due to the existence of (i) abundant and complex evolving interfaces, such as material interface and mechanical interfaces associated with substantial gradient forces; (ii) complicated multi-scale spatio-temporal structures and their cross-scale correlations, such as vortex, core center, braid, growing, collapsing, deformation, breaking up, even turbulence; (iii) competition between various scales and kinetic modes. Therefore, besides the hydrodynamic nonequilibrium (HNE) characteristics, accurate modeling and understanding the cross-scale process requires to carefully take into account the TNE effects, which not only help to dynamically characterize the nonequilibrium state but also refine the constitutive relations.

\emph{Second, the effect of heat conduction or ablation on KHI needs further investigation.}
It is investigated limitedly and the conclusion is highly controversial. On the one hand, Refs. \cite{49,50} suggest
that heat conduction stabilizes the KHI through reducing the linear growth rate and frequency, suppressing the transmission of the perturbation and the appearance of higher-order harmonics; while strengthening the pairing and the formation of large-scale structures for a two-mode sinusoidal interface perturbation. On the other hand, in Refs. \cite{51,52,53,54}, viscous potential flow analysis on KHI between liquid and vapor phases of a fluid
indicate that heat transfer has a destabilizing influence on the relative
velocity and the stability of the system.

\emph{Third, the understanding of complex fields resulted from KHI is far from clear.} It is well-known that a large variety of complex spatial patterns spring up with the evolution of KHI. How to effectively describe the pattern dynamics and pick up more information from such a complicated system is still an open problem.

In this paper, we would like to address the three aspects described above.
Specifically, through modifying the collision term of the discrete Boltzmann
equation, we propose an efficient and easily implementable
discrete Boltzmann model (DBM) for two-dimensional compressible flows with
flexible specific-heat ratio and Prandtl number.
Next, we introduce the Minkowski functionals, which are well known in
digital-picture analysis \cite{78}, and has been successfully
applied to characterize patterns in phase-separating system \cite{79,80,81}, shocked porous materials \cite{59,82}, etc, to
extract information from the complex patterns emerging from the evolution of
KHI. Finally, via the DBM simulation and the Minkowski measures, we focus on
the TNE and morphological behaviors, and aim to clarify the viscosity and
heat conduction effects on the onset and growth of KHI, quantitatively.

The rest of the paper is structured as follows.
In section 2, we briefly review the DBM used in this work, then verify and validate the model through three typical test cases in section 3.
Effects of viscosity and heat conduction on
KHI are studied in detail in section 4. Section 5 summarizes and concludes the present paper.

\section{DBM with flexible specific-heat ratio and Prandtl number}

 The model we presented here
consists of two components: (i) a highly
efficient DBM for compressible flows with flexible specific-heat ratio \cite{60}; (ii) a modification to the BGK collision term for achieving flexible Prandtl number.

\subsection{Highly efficient DBM with flexible specific-heat ratio}

The foremost essence of the DB modeling is the construction of the discrete
equilibrium distribution function (DEDF) ${f}_{i}^{(0)}$ which decides the
physical accuracy, numerical stability and efficiency of DBM \cite{60,83}. In 2013, we
presented a simple and general approach to formulate DEDF through inversely
solving the kinetic moments that DEDF satisfies \cite{60}. The
crucial physical requirement is that all the required kinetic moments of ${f}%
_{i}^{(0)}$ in the discrete summation form, should
equal to those in the integral form of ${f}^{(0)}$,
\begin{equation}
\sum\nolimits_{i}{f_{i}^{(0)}}\Psi _{i}{(}\mathbf{v}_{i},\eta _{i})={\mathbf{M}}_{m,n}={\iint {{{f}^{(0)}}}}\Psi {(}\mathbf{v,}\eta {)}d\mathbf{v}d\eta
\text{,}  \label{Moment}
\end{equation}%
where
${{f}^{(0)}}={{\frac{{\rho }}{2\pi T}%
}}{{\ \left( \frac{{1}}{2\pi b T}\right) }^{1/2}}\exp \left[ -\frac{{{\left(\mathbf{v}-\mathbf{u}\right) }^{2}}}{2T}-\frac{{{\eta }^{2}}}{2 b T}\right] $,
$\Psi _{i}(\mathbf{v}_{i},\eta _{i})=[1,\mathbf{v}_{i},\frac{1}{2}%
(v_{i}^{2}+\eta _{i}^{2}),...]^{T}$, with
$\rho$, $\mathbf{v}$, $\mathbf{u}$, $T$ are the local density, particle velocity,  flow velocity and temperature, respectively.
$\mathbf{v}_{i}$ is the discrete-velocity model (DVM),
$\eta _{i}$ is a set of free parameters introduced to describe the $b$ extra degrees of freedom corresponding to
molecular rotation and/or vibration.
So the specific-heat ratio $\gamma
=(b+4)/(b+2)$. $\mathbf{M}_{m,n}$ means that the $m$-th tensor is contracted
to a $n$-th one. Chapman-Enskog analysis demonstrates that, to recover the
thermohydrodynamic equations at the Navier-Stokes level, it needs $\Psi _{i-\text{NS}}(\mathbf{v}_{i},\eta _{i})=[1,\mathbf{v}_{i},\frac{1}{2}%
(v_{i}^{2}+\eta _{i}^{2}),\mathbf{v}_{i}\mathbf{v}_{i},\frac{1}{2}%
(v_{i}^{2}+\eta _{i}^{2})\mathbf{v}_{i},\mathbf{v}_{i}\mathbf{v}_{i}\mathbf{v%
}_{i},\frac{1}{2}(v_{i}^{2}+\eta _{i}^{2})\mathbf{v}_{i}\mathbf{v}_{i}]^{T}$
with $i=1,2,...,16$ for two-dimensional case. We rewrite Eq. (\ref{Moment})
in a matrix form
\begin{equation}
{\bm{\Psi }}_{\text{NS}}\cdot \mathbf{f}^{(0)}={\mathbf{M}}_{m,n}\text{,}
\label{Feq_1}
\end{equation}%
then $\mathbf{f}^{(0)}$ can be formulated as
\begin{equation}
\mathbf{f}^{(0)}={\bm{\Psi }}_{\text{NS}}^{-1}\cdot {\mathbf{M}}_{m,n}%
\text{,}
\end{equation}%
with ${\bm{\Psi }}_{\text{NS}}^{-1}$ the inverse of matrix ${\bm{\Psi }}_{\text{NS}}=(\Psi _{1-\text{NS}},\Psi _{2-\text{NS}},\cdots ,\Psi
_{16-\text{NS}})^{T}$, $\mathbf{f}^{(0)}=(f_{1}^{(0)},f_{2}^{(0)},\cdots
,f_{16}^{(0)})^{T}$, ${\mathbf{M}}%
_{m,n}=(M_{0},M_{1x},M_{1y},...,M_{4,2yy})^{T}$ is the set of moments of $
f_{i}^{(0)}$. The following two-dimensional DVM with $16$ discrete
velocities has been designed to discretize the velocity space and to ensure
the existence of ${\bm{\Psi }}_{\text{NS}}^{-1}$ \cite{60}
\begin{equation}
(v_{ix},v_{iy})=\left\{
\begin{array}{cc}
\text{cyc}:c(\pm 1,0) & \text{for}\quad 1\leq i\leq 4 \\
c(\pm 1,\pm 1) & \text{for}\quad 5\leq i\leq 8 \\
\text{cyc}:2c(\pm 1,0) & \ \text{for}\quad 9\leq i\leq 12 \\
2c(\pm 1,\pm 1) & \ \quad \text{for}\quad 13\leq i\leq 16%
\end{array}%
\right. ,
\end{equation}%
where \textquotedblleft cyc" represents the cyclic permutation. For $1\leq
i\leq 4$, $\eta _{i}=\eta _{0}$; otherwise, $\eta _{i}=0$. Here $c$ and $%
\eta _{0}$ are two free parameters, adjusted to optimize the property of
the model.

For discrete Boltzmann modeling, the moment system that we aim to match is the extended Maxwell-Boltzmann moment system which owns flexible specific-heat ratio. Similar to the Machado moment system \cite{83}, it is also a system rather than few moments.
For a specific model, say the the discrete Boltzmann model we presented here, ${f_{i}^{(0)}}$ only satisfies 7 kinetic moments which are exactly consistent with the Kataoka-Tsutahara moments \cite{84}.
To access behaviors of the system farther-away-from equilibrium \cite{85,86,87},
besides the 7 kinetic moments, ${f_{i}^{(0)}}$ should satisfy more nonhydrodynamic kinetic moments. Consequently, ${\Psi}(\mathbf{v}_{i},\eta _{i})$ owns more elements,  $\mathbf{v}_{i}$ and $\eta_{i}$ own more discrete velocities, ${f_{i}^{(0)}}$ becomes more complex, and naturally, the DBM owns more powerful multi-scale predictive capability.
Compared with the corresponding
hydrodynamic equations, e.g., Burnett or Super-Burnett equations, whose complexity increases substantially with increasing the degree of TNE effects,
the complexity of DBM modeling increases negligibly \cite
{56,71}.

\subsection{Modification to the BGK collision term }

The BGK discrete Boltzmann equation utilizes a single relaxation time in the
collision term which results in a fixed Prandtl number $\Pr =1$.
To overcome this limitation, several strategies within different frameworks have been developed.
The first is the replacement of distribution function approach, where the
original Maxwellian has been substituted by an anisotropic Guassian distribution. Models belonging to this category include the ellipsoidal statistical BGK model (modify the stress tensor) \cite{88}, Shakhov model (modify the heat flux) \cite{89}, the Liu model (modify both the stress tensor and heat flux) \cite{90}, etc.
The second
is the two-relaxation-time or  \cite
{91} the
multiple-relaxation-time approach \cite
{63,92,93}, where more free parameters are
introduced to describe the relaxation rates of various kinetic moments due to particle collisions, then results in the flexible Prandtl number.
Besides, Machado contributed a more general moment system which has an intrinsic extra $\mu$ term \cite{94}.
A flexible Prandtl number can be obtained in both the Boltzmann equation
and in the lattice Boltzmann equation under the BGK collision framework through changing the targeted local equilibrium state via the $\mu$ term.
Here we also present one way to adjust the Prandtl number under the BGK framework through adding an external forcing term $\Theta _{i}$ into the right-hand-side of the discrete Boltzmann equation to modify the BGK collision operator,
\begin{equation}
\frac{\partial f_{i}}{\partial t}+\mathbf{v}_{i}\cdot \bm{\nabla}f_{i}=-%
\frac{1}{\tau }[f_{i}-f_{i}^{(0)}]+\Theta _{i},  \label{DB equation}
\end{equation}%
where $\Theta _{i}=-B[2T-(\mathbf{v}_{i}-\mathbf{u})^{2}]f_{i}^{(0)}$ with $%
B=\frac{1}{2\rho T^{2}}\bm{\nabla}\cdot (\frac{4+b}{2}\rho T\theta %
\bm{\nabla}T)$. Correspondingly, the heat conductivity has been changed to
be $\kappa _{T}=\frac{4+b}{2}\rho T(\tau +\theta )$, and the Prandtl number $%
\Pr =\tau /(\tau +\theta )$. One of the prominent advantages is that this
modification does not give rise to additional kinetic moment requirements on
$f_{i}^{(0)}$. After that, we solve Eq. (\ref{DB equation}) to update $f_{i}$
via numerical schemes. Hydrodynamic quantities, such as density, momentum,
total energy can be obtained from kinetic moments of $f_{i}$, $\rho
=\sum\nolimits_{i}f_{i}$, $\rho \mathbf{u}=\sum\nolimits_{i}f_{i}\mathbf{v}%
_{i}$ and $E=\rho (c_{v}T+\frac{1}{2}u^{2})=\sum\nolimits_{i}\frac{1}{2}%
f_{i}(v_{i}^{2}+\eta _{i}^{2})$ with $c_{v}=\frac{2+b}{2}$ the specific heat
at constant volume. The pressure can be calculated from the equation of
state for ideal gases $P=\rho T$.

\section{Verification and validation}

In this section, several typical benchmarks, including the thermal plane Couette flow problem \cite{95}, the Sod shock tube problem \cite{96} and the Modified Colella's explosion wave test case \cite{97}, ranging from subsonic to supersonic, have been conducted to validate the new model.
The discrete Boltzmann equation, particle velocity, and hydrodynamic quantities have been nondimensionalized by suitable reference variables as listed in Ref. \cite{56}.

To ensure numerical stability and accuracy, the fifth-order weighted essentially nonoscillatory (5th-WENO) finite
difference (FD) scheme \cite{98} is employed to discretize the spatial derivatives for the first two test cases and the latter KHI simulations; the second-order non-oscillatory non-free-parameter and dissipative (NND) FD scheme \cite{99} is used to discretize the spatial derivatives for the third
Riemann problem; the second-order implicit-explicit Runge-Kutta FD scheme
\cite{100} is utilized to solve the temporal derivative for all test cases.
Compared with the standard lattice Boltzmann method where particle velocities are restricted to fixed values and exactly link the lattice sites in unit time,
the utilization of the FD scheme gets rid of the banding of spatial and temporal discretizations.
The sets of particle velocities enjoy high degrees of freedom in configuration, magnitude and number.
Consequently, they are much more convenient to meet the stability, robustness and accuracy requirements for simulating compressible nonequilibrium flows with Mach number as high as 30.
Of course, the adoption of the FD scheme will inevitably introduce numerical errors and
make the implementation of boundary conditions (BCs)  incorporated into the model intricately.
For example, when using the elaborate 5th-WENO algorithm,
three ghost nodes out of the boundary are needed at each side of the boundary
in the presence of solid walls.
Details on how to implement BCs with this scheme can be found in Ref. \cite{47}.
As for numerical errors, they decrease sharply with
finer mesh size and smaller time step when the above mentioned total variation diminishing schemes have been applied. So in our simulations, the finest mesh
and small enough time step,
which the model and the computational resource can undergo respectively, are employed for each test.
At last, we suggest that, the 5th-WENO scheme is more effective in decreasing the numerical dissipation and improving the accuracy compared to the NND scheme,
since it changes the method of choosing smooth stencil with logical judgment into weighted average of all stencils.
But the NND scheme is more stable than the former owing to its slightly stronger dissipation, especially in regions near discontinuities.
Therefore, for test cases without shock wave or with weak shock wave, the 5th-WENO scheme is preferred; for cases containing strong shock, the NND scheme is preferred.

\subsection{Thermal plane Couette flow}

%%%%%%%%%%%%%%%%%%%%%%%%%%%%%%%%%%%%%%%%%%%%%%%%%%%%%%%%%%%%%%%%%%%%%%%%%%%%%%%%%
\begin{figure*}[b!]
	\centering
	\includegraphics[width=0.9\textwidth]{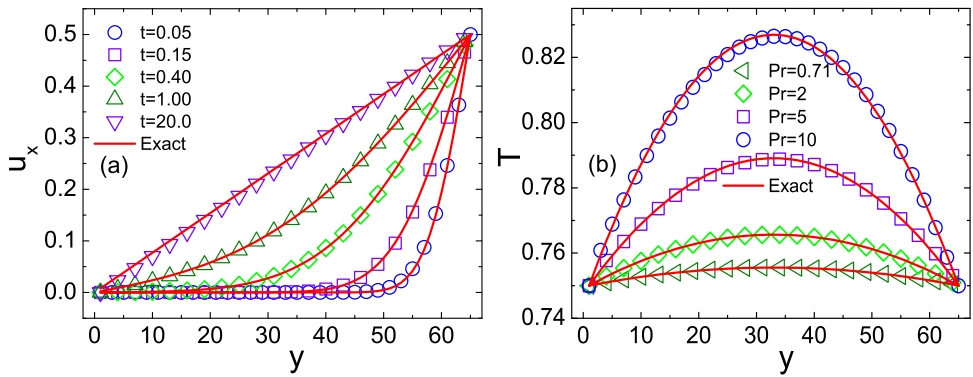}
	\caption{{\bf(a)} Velocity profiles in unsteady Couette flow at characteristic times. 
		{\bf(b)} Temperature profiles in steady Couette flow for cases with various
		Prandtl numbers.}
\end{figure*}
%%%%%%%%%%%%%%%%%%%%%%%%%%%%%%%%%%%%%%%%%%%%%%%%%%%%%%%%%%%%%%%%%%%%%%%%%%%%%

As a classical test where viscosity and heat transfer dominate, the thermal plane
Couette flow is commonly applied to examine the ability of the DBM for
simulating compressible flows with flexible Prandtl number \cite{71,101,102}. For this problem
considered, a viscous fluid flow between two infinite parallel flat plates
separated by a constant distance of $H$, possesses the following initial
conditions $(\rho ,T,u_{x},u_{y})=(1.0,0.75,0.0,0.0)$. When the simulation
starts, the top plate moves horizontally at the speed of $u_{0}=0.5$, while
the bottom plate keeps stationary. Periodic BCs and
nonequilibrium extrapolation scheme are applied in the $x$ and $y$
directions, respectively. Simulations have been carried out on a uniform
mesh $N_{x}\times N_{y}=4\times 65$ with $\Delta x=\Delta y=3\times 10^{-3}$%
, $\tau =100\Delta t=10^{-3}$, $c=0.84$ and $\eta _{0}=2.0$. Figure 1(a)
shows comparisons between the distributions of the horizontal velocity $%
u_{x} $ along the $y$-axis and the exact solutions at characteristic times $%
t=0.05$, $0.15$, $0.40$, $1.00$, $20.0$, respectively. To be noticed is that
the simulation results agree well with the analytical solutions
\begin{equation}
u_{x}=\frac{y}{H}u_{0}+\frac{2}{\pi }u_{0}\overset{\infty }{\underset{n=1}{%
		\sum }}[\frac{(-1)^{n}}{n}\exp (-n^{2}\pi ^{2}\frac{\mu t}{\rho H^{2}})\sin (%
\frac{n\pi y}{H})]\text{,}
\end{equation}%
with $\mu =P\tau $ the viscosity coefficient. Figure 1(b) displays
temperature profiles along the $y$-direction in the steady state for cases
with different Prandtl numbers, where the following theoretical solutions
are also exhibited for comparisons
\begin{equation}
T=T_{0}+\frac{\Pr }{2c_{p}}u_{0}^{2}\frac{y}{H}(1-\frac{y}{H})\text{.}
\end{equation}%
Here $T_{0}=0.75$ is the temperature of the top/bottom wall, $c_{p}=\gamma
c_{v}$ is the specific heat at constant pressure with $\gamma =4/3$. As
shown, the simulation results also match well with the corresponding
analytical solutions, indicating the validity of the DBM in mimicking
compressible flows with flexible Prandtl number.

\subsection{Sod shock tube}

%%%%%%%%%%%%%%%%%%%%%%%%%%%%%%%%%%%%%%%%%%%%%%%%%%%%%%%%%%%%%%%%%%%%%%%%%%%%%%%%%
\begin{figure*}[b!]
	\centering
	\includegraphics[width=0.88\textwidth]{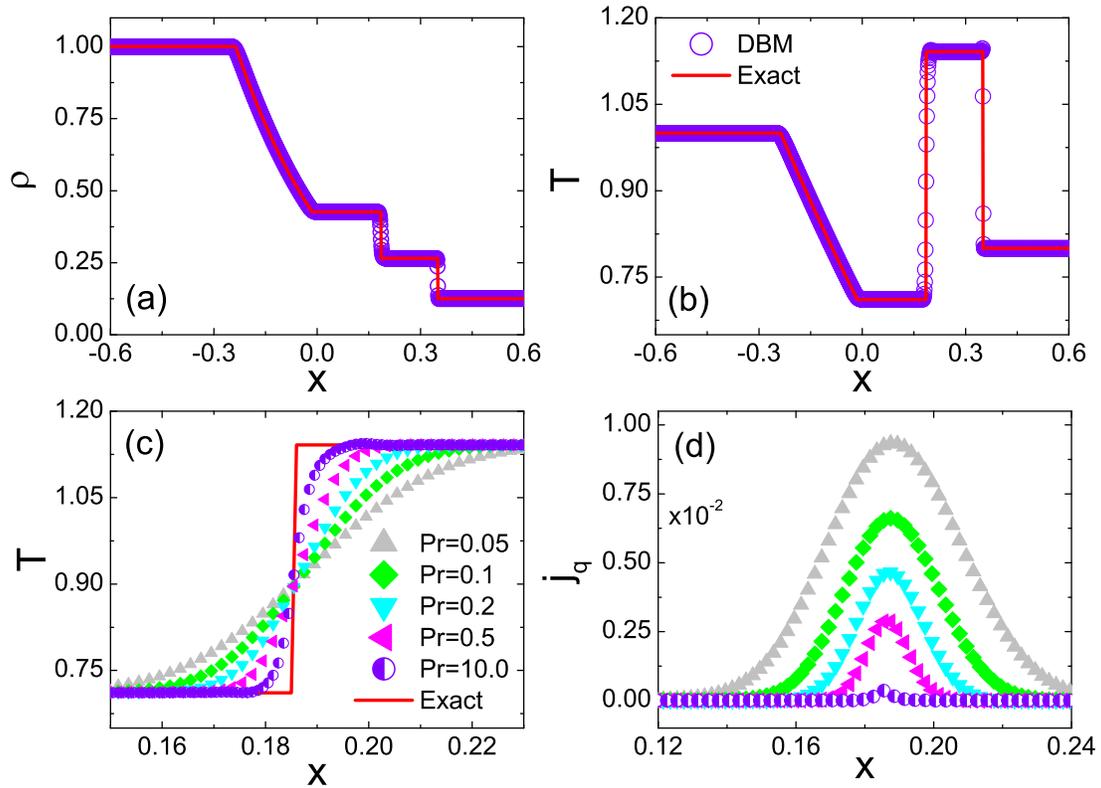}
	\caption{ Profiles of density {\bf(a)}, temperature {\bf(b)},  local
		temperature {\bf(c)} and heat flux {\bf(d)} nearby the contact
		discontinuity with various Prandtl numbers calculated from DBM and Riemann
		solutions for the Sod shock tube, where $t=0.2$ and $\gamma=1.4$.}
\end{figure*}
%%%%%%%%%%%%%%%%%%%%%%%%%%%%%%%%%%%%%%%%%%%%%%%%%%%%%%%%%%%%%%%%%%%%%%%%%%%%%%%%%

Due to the inclusion of rich and complex characteristic structures, the Sod
shock tube problem is also a classical test used to verify the performance
of models for compressible flow. The initial conditions are
\begin{equation}
\left\{
\begin{array}{c}
(\rho ,T,u_{x},u_{y})|_{L}=(1.0,1.0,0.0,0.0), \\
(\rho ,T,u_{x},u_{y})|_{R}=(0.125,0.8,0.0,0.0),%
\end{array}%
\right.
\end{equation}%
where subscripts \textquotedblleft L" and \textquotedblleft R" stand for
macroscopic variables at the left and right sides of the discontinuity,
respectively. In the $x$ and $y$ directions, we adopt the supersonic inflow and  periodic BCs, respectively. Parameters are set to be $\Delta
x=\Delta y=10^{-3}$, $\Delta t=5\times 10^{-5}$, $c=1.06$, $\eta _{0}=1.5$, $%
\gamma =1.4$, $N_{x}\times N_{y}=2000\times 4$. The relaxation time $\tau $
is fixed at $5\times 10^{-5}$ for all simulations. Shown in Fig. 2 are the
computed profiles of density, temperature, local temperature and heat flux
nearby the contact discontinuity for cases with various Prandtl numbers at $%
t=0.2$, where solid lines indicate Riemann solutions. From the top two
subgraphs, it is clear that the left-propagating rarefaction wave, the
right-propagating shock wave and the contact discontinuity have been exactly
captured with severely curtailed numerical dissipation. Enlargement of the
local part containing shock wave manifests that the shock wave only spreads
over three to four grid cells. For a fixed viscosity, the increase in
Prandtl number leads to the decrease in heat conductivity. Figures 2(c)-(d)
inform us that heat conduction smoothes the temperature profile,
reduces the temperature gradient, but enlarges the amplitude of heat flux
and extends the nonequilibrium region.

\subsection{Modified Colella's explosion wave test}

%%%%%%%%%%%%%%%%%%%%%%%%%%%%%%%%%%%%%%%%%%%%%%%%%%%%%%%%%%%%%%%%%%%%%%%%%%%%%%%%%
\begin{figure*}[b!]
	\centering
	\includegraphics[width=0.88\textwidth]{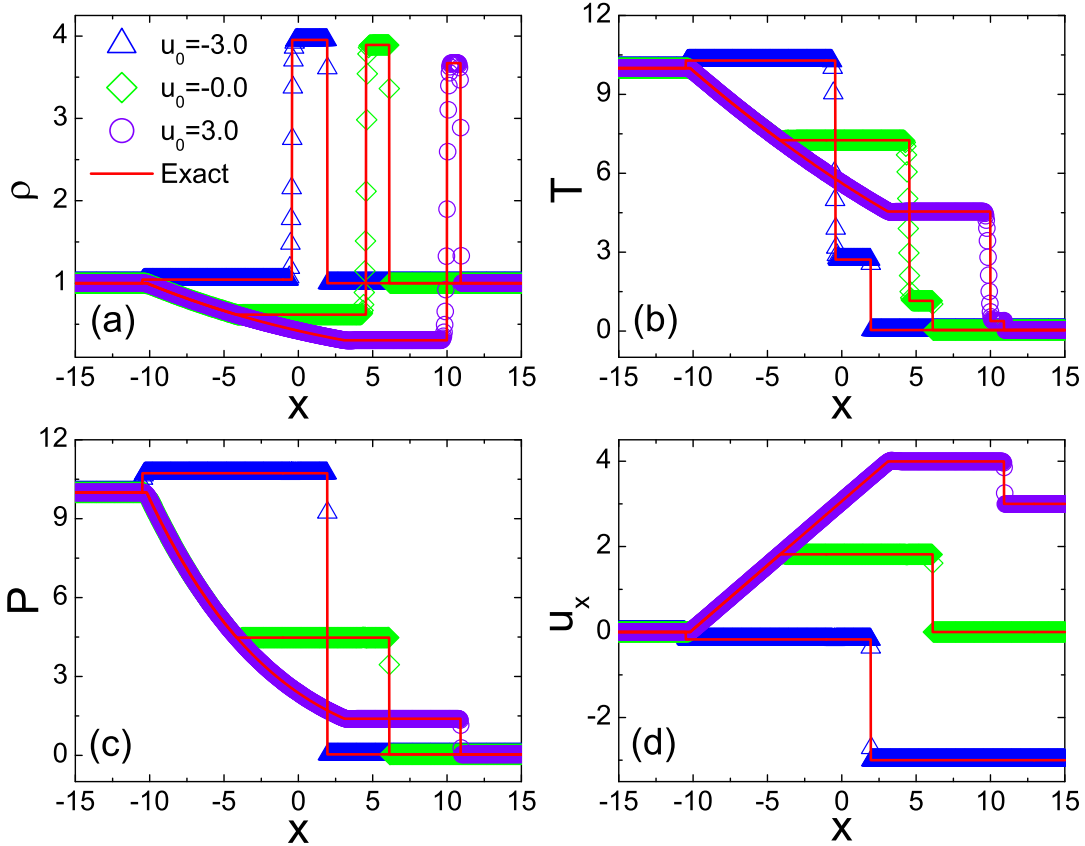}
	\caption{ Profiles of density {\bf(a)}, temperature {\bf(b)},  pressure
		{\bf(c)} and velocity {\bf(d)} with various initial conditions
		calculated from DBM and Riemann solutions for the modified Colella's shock
		tube, where $t=2.5$, $\gamma=5/3$ and $\Pr=0.71$.}
\end{figure*}
%%%%%%%%%%%%%%%%%%%%%%%%%%%%%%%%%%%%%%%%%%%%%%%%%%%%%%%%%%%%%%%%%%%%%%%%%%%%%%%%%

To further examine the robustness and effectiveness of the model for
compressible flows with high Mach number, we construct modified Colella's
explosion wave tests with weaker temperature discontinuity but stronger
velocity discontinuity
\begin{equation}
\left\{
\begin{array}{c}
(\rho,T,u_{x},u_{y})|_{L}=(1.0,10.0,0.0,0.0), \\
(\rho,T,u_{x},u_{y})|_{R}=(1.0,1/30,u_{0},0.0).%
\end{array}%
\right.
\end{equation}
Comparisons between simulation results and the exact solutions at $t=2.5$
are plotted in Fig. 3, with $u_{0}=-3.0$, $0.0$, and $3.0$. Parameters used
here are $\Delta x=\Delta y=10^{-2}$, $\Delta t=5\times 10^{-5}$, $\tau
=2\times 10^{-5}$, $c=2.06$, $\eta _{0}=20.0$, $\gamma =5/3$, $N_{x}\times
N_{y}=4000\times 2$. The simulation results agree excellently with Riemann
solutions for each case. Moreover, the shock wave and contact discontinuity
are captured stably without overshoots or spurious oscillations. Successful
simulation of this aggressive tests manifests that the proposed model is
robust, accurate and applicable to compressible flows with high-Mach-number (
$Ma=12.7$ for case with $u_{0}=3.0$), high temperature and pressure ratios
(up to $300$).

\section{Effects of viscosity and heat conduction on KHI}

In this section, we conduct a parametric study to evaluate the effects of
viscosity and heat conduction on the formation and evolution of the KHI.
Both the HNE and TNE manifestations provided by
DBM, as well as morphological characterizations described by
Minkowski measures, have been extracted to analyze and understand the
complex configurations and nonequilibrium
processes. For all simulations, the whole two dimensional calculation domain
corresponds to a rectangle with length $L_{x}=1.8$ and height $L_{y}=0.6$,
divided into $600\times 200$ grid cells.

%%%%%%%%%%%%%%%%%%%%%%%%%%%%%%%%%%%%%%%%%%%%%%%%%%%%%%%%%%%%%%%%%%%%%%%%%%%%%%%%%
\begin{figure*}[b!]
	\centering
	\includegraphics[width=0.88\textwidth]{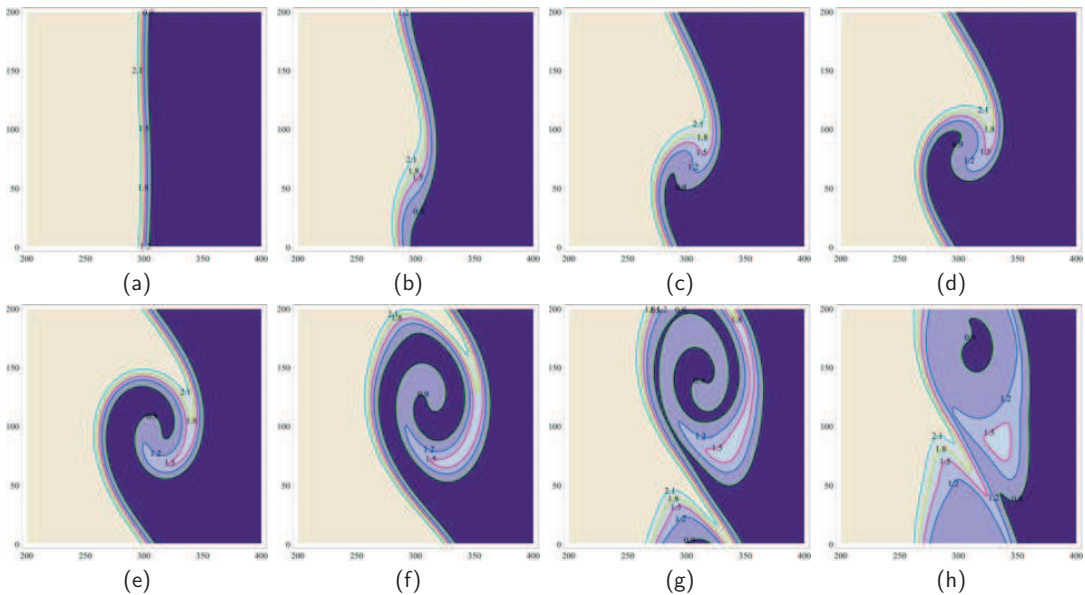}
	\caption{ Time series of the density patterns simulated via the DBM, where $t=0.1$ in {\bf(a)}, $t=0.7$ in {\bf(b)}, $t=1.0$ in {\bf(c)}, $t=1.2$ in {\bf(d)}, $t=1.5$ in
		{\bf(e)}, $t=2.0$ in {\bf(f)}, $t=2.45$ in {\bf(g)}, and $t=2.85$ in {\bf(h)}. Only the part
		with $x$ in the range $[200:400]$ of the full grid is shown. From purple to light brown, the density increases.}
\end{figure*}
%%%%%%%%%%%%%%%%%%%%%%%%%%%%%%%%%%%%%%%%%%%%%%%%%%%%%%%%%%%%%%%%%%%%%%%%%%%%%%%%%

\subsection{Density patterns, nonequilibrium and morphological characterizations}

The initial configurations of our simulations are functions of $x$, read
\begin{equation}
\rho (x)=\frac{{\rho _{L}+\rho _{R}}}{2}-\frac{{\rho _{L}-\rho _{R}}}{2}%
\tanh (\frac{x-L_x/2}{{D_{\rho }}})\text{,}
\end{equation}%
\begin{equation}
u_{y}(x)=\frac{{u_{yL}+u_{yR}}}{2}-\frac{{u_{yL}-u_{yR}}}{2}\tanh (\frac{x-L_x/2}{{D_{u}}})\text{,}
\end{equation}%
\begin{equation}
P_{L}=P_{R}=P\text{,}
\end{equation}%
where $D_{\rho }=4$ ($D_{u}=4$) indicates the width of density (velocity)
transition layer.
$\rho _{L}=2.4$ ($\rho _{R}=0.6$) is the density away from the interface of
the left (right) fluid. $u_{yL}=-u_{yR}=0.5$ stands for the fluid owning
opposite vertical velocities in the two halves, while having homogeneous
pressure $P=1.2$.
In the absence of any perturbation, the configuration maintains in
mechanical equilibrium. To trigger the KH rollup, we introduce a small
velocity perturbation in the $x$ direction as
\begin{equation}
u_{x}(x)=u_{0}\sin (ky)\exp (-2\pi |x-L_{x}/2|)\text{,}  \label{uu}
\end{equation}%
where $u_{0}=20\Delta x$ denotes the amplitude of the initial perturbation
and $k=2\pi /L_{y}$ is the wave number of the initial perturbation. Periodic BCs are
applied in the $y$ direction and outflow boundary conditions are adopted in
the $x$ direction. The time step is set to be as small as $\Delta t=10^{-5}$
to reduce the numerical dissipation. The remaining parameters are $c=1.2$, $%
\eta _{0}=200$, $\tau =2\times 10^{-4}$ and $\gamma =5/3$.

%%%%%%%%%%%%%%%%%%%%%%%%%%%%%%%%%%%%%%%%%%%%%%%%%%%%%%%%%%%%%%%%%%%%%%%%%%%%%%%%%
\begin{figure*}[b!]
	\centering
	\includegraphics[width=0.85\textwidth]{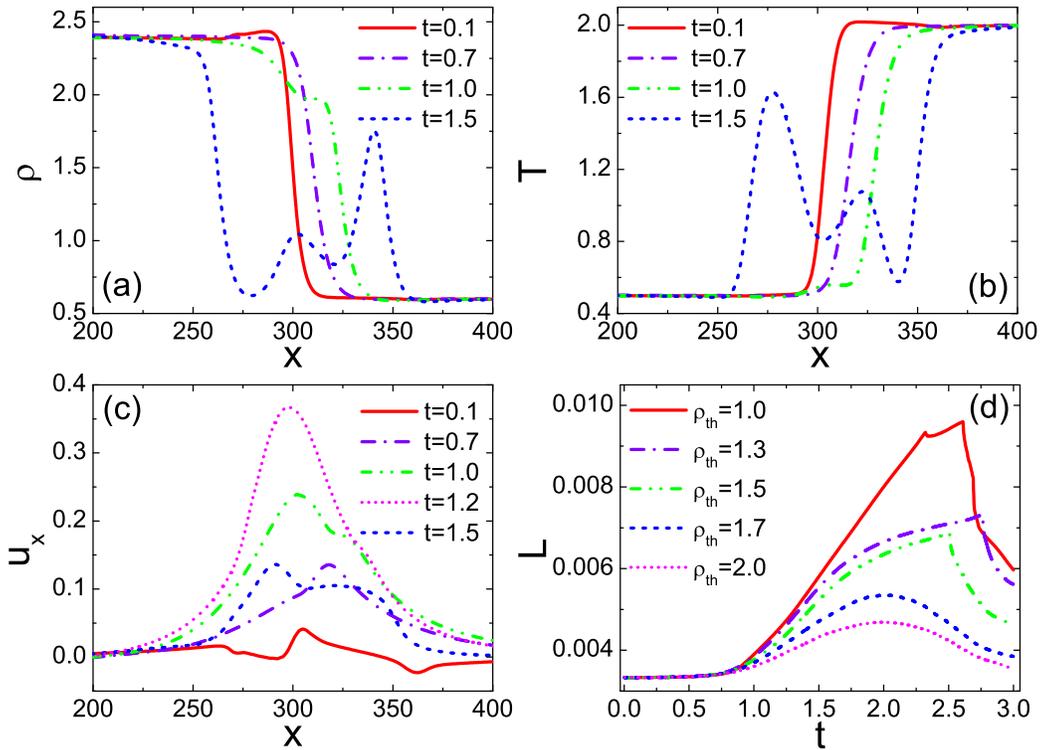}
	\caption{Profiles of density {\bf(a)}, temperature {\bf(b)} and
		velocity{\bf(c)} along the horizontal centerline $y=L_y/2$ at various
		times, and time evolutions of boundary length $L$ {\bf(d)} for various
		density thresholds $\rho_{\text{th}}$.}
\end{figure*}
%%%%%%%%%%%%%%%%%%%%%%%%%%%%%%%%%%%%%%%%%%%%%%%%%%%%%%%%%%%%%%%%%%%%%%%%%%%%%%%%%

Figure 4 shows time series of density patterns with five contour lines
during the evolution of KHI. From it, four distinct evolutionary regimes,
i.e., the oscillatory regime, the linear growth stage, the nonlinear growth
stage with highly rolled-up vortices, and finally, a turbulent phase with
nonregular structures, can be
distinguished.
Specifically, in the first stage [see panel (a)], the continuous interface
between the two layers begins wiggling under the action of the initial
perturbation and the velocity shear. After such a transient period, the
perturbation grows exponentially and results in a sinuous structure
dominating in panels (b)-(d). In the subsequent nonlinear phase, a roughly
circular vortex is formed at the expense of the vortex in the braid regions
[see panel (e)]. After saturation, the vortex is further stretched in the $y$
direction and becomes elliptical as shown in panel (f). In the final stage,
the normal vortex structure collapses and the mixing layer approaches the
boundary, marking that the system proceeds to the turbulent stage. Moreover,
it is observed that the position of the mixing layer moves toward the $x$
direction from the center of the computational domain, which enhances the
transfer of fluids from the dense to the tenuous region.

To further understand the development of the vortex or the mixing layer, in
Fig. 5, we illustrate the density, temperature and velocity profiles along
the horizontal centerline $y=L_y/2$ at representative times.
The density and temperature profiles vary from being smooth (tangent
profiles) to being irregular. The width of the mixing layer and the
amplitude of the oscillation increase with time, due to the
mass-momentum-energy transport from the dense (hot) to the rarefactive
(cold) regions, see the crest at $x=340$ ($x=277$) in the density
(temperature) profile at $t=1.5$ and the fully developed horizontal velocity
$u_x(x)$, which attains a maximum at $t=1.2$ (roughly $70\%$ of $u_{yL}$),
before decreasing monotonically to zero.

%%%%%%%%%%%%%%%%%%%%%%%%%%%%%%%%%%%%%%%%%%%%%%%%%%%%%%%%%%%%%%%%%%%%%%%%%%%%%%%%%
\begin{figure*}[b!]
	\centering
	\includegraphics[width=0.92\textwidth]{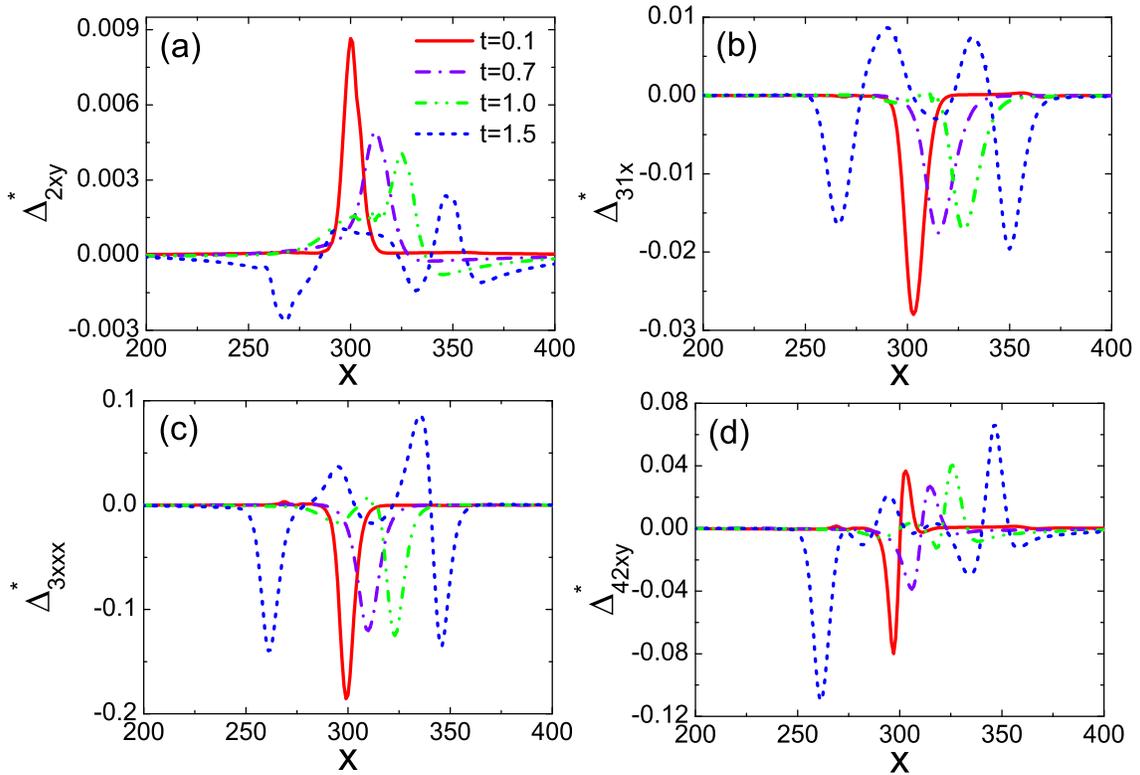}
	\caption{Nonequilibrium measures $\Delta_{2xy}^*$ {\bf(a)}, $
		\Delta_{31x}^*$ {\bf(b)}, $\Delta_{3xxx}^*$ {\bf(c)}, and $\Delta_{42xy}^*$ {\bf(d)}
		along the horizontal centerline $y=L_y/2$ at various times.}
\end{figure*}
%%%%%%%%%%%%%%%%%%%%%%%%%%%%%%%%%%%%%%%%%%%%%%%%%%%%%%%%%%%%%%%%%%%%%%%%%%%%%%%%%

Next, the Minkowski measures are presented to extract information from the
complex patterns displayed in Fig. 4. According to the morphological
analysis, a physical field $\Theta(\mathbf{r},t)$ of interest can be
condensed as two kinds of characteristic regimes: the white (with $%
\Theta>\Theta_{\text{th}}$) and the black (with $\Theta<\Theta_{\text{th}}$%
), where $\Theta_{\text{th}}$ is a threshold of $\Theta$. For such a Turing
pattern, a general theorem of integral geometry states that all the
properties of a $d$-dimensional convex set satisfying motion invariance and
additivity are contained in $d+1$ Minkowski measures. To be specific, for a
two dimensional density field, the three Minkowski measures are the total
fractional area $A$ of the high-density regimes, the boundary length $L$
between the high- and low-density regimes per unit area, and the Euler
characteristic $\chi$ per unit area. Figure 5(d) depicts the
time evolutions of boundary length $L$ for various density thresholds $\rho_{%
	\text{th}}$, in a log-linear scale. As shown, these curves behave
qualitatively similar and can be roughly divided into four stages, marked
individually by red arrows for $L(t)$ curve with $\rho_{\text{th}}$ equaling
to the averaged density. The first stage
corresponds to the time delay for observing evident KH billows. Still no
notable instability occurs for all the density thresholds. Afterwards, $L$
increases %sharply
exponentially till $t=1.5$, followed by a slow increase to attain the peak
at about $t=2.5$;
whereafter decreases abruptly, especially for cases with smaller thresholds.
The first increase and the subsequent decrease in $L$ are due to the growth,
formation, and deformation of the KH rolls from small fluctuations
undergoing
the linear and nonlinear stages, and the finally coalesce of the high- and
low-density domains by the secondary instabilities that induces vortex
breakup during the oversaturated turbulent stage,
respectively.

%%%%%%%%%%%%%%%%%%%%%%%%%%%%%%%%%%%%%%%%%%%%%%%%%%%%%%%%%%%%%%%%%%%%%%%%%%%%%%%%%
\begin{figure*}[b!]
	\centering
	\includegraphics[width=0.65\textwidth]{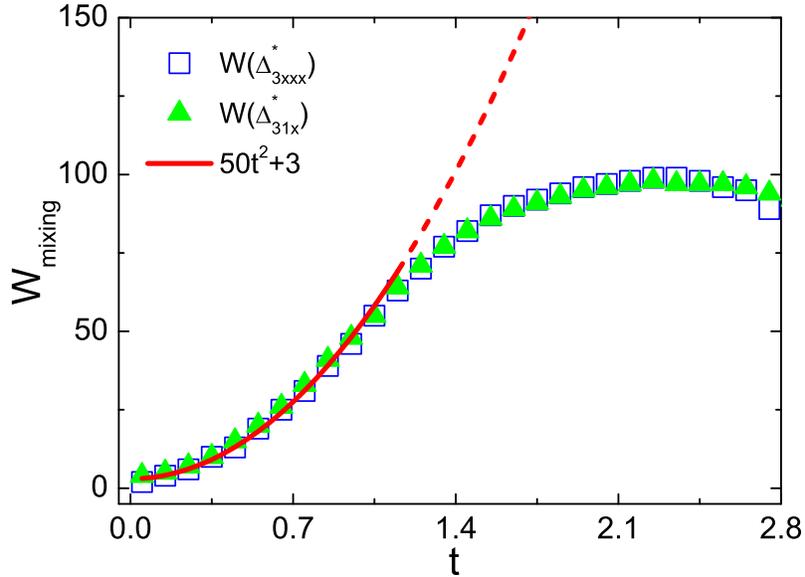}
	\caption{Widths of the mixing layers obtained from
		interface-tracking technique based on the nonequilibrium measures
		$\Delta_{3xxx}^*$ and $\Delta_{31x}^*$.}
\end{figure*}
%%%%%%%%%%%%%%%%%%%%%%%%%%%%%%%%%%%%%%%%%%%%%%%%%%%%%%%%%%%%%%%%%%%%%%%%%%%%%%%%

Besides being able to recover hydrodynamic equations at various levels (Navier-Stokes,
Burnett, super-Burnett, etc.), DBM also provides us a set of handy,
effective and efficient tools to describe, measure, and analyze the TNE
behaviors, by calculating the difference between the non-conserved central
kinetic moments of discrete distribution function and DEDF, $\bm{\Delta}%
_{m,n}^{\ast }=\mathbf{M}_{m,n}^{\ast }(f_{i})-\mathbf{M}_{m,n}^{\ast
}(f_{i}^{(0)})=\sum\nolimits_{i}(f_{i}-f_{i}^{(0)})(\frac{v_{i}^{\ast
		2}+\eta _{i}^{2}}{2})^{\frac{m-n}{2}}\mathbf{v}_{i}^{\ast n}$,
% where ``$m,n$" means that the $m$-th tensor is contracted to a $n$-th one,
with $\mathbf{v}_{i}^{\ast }=\mathbf{v}_{i}-\mathbf{u}$ the thermal velocity.
Figure 6 qualitatively portrays the typical nonequilibrium manifestations $%
\Delta _{2xy}^{\ast }$, $\Delta _{3,1x}^{\ast }$, $\Delta _{3xxx}^{\ast }$,
and $\Delta _{4,2xy}^{\ast }$ during the development of KHI.
The following features can be obtained: (i) For an ideal gas system, gradient force acts as
the unique driving force for TNE and HNE. So,
the TNE quantities are mostly around the interface where the gradients of
macroscopic quantities are pronounced and exactly attain their local maxima
(minima) at the points of the maxima ($\bm{\nabla}\rho ,\bm{\nabla}T,%
\bm{\nabla}\bm{u})_{\text{max}}$; while approach zero at positions far away
from the interface or at peaks (valleys) with vanishing gradients. For
example, at the peak ($x=277$) and valley ($x=340$) of $T(x,L_{y}/2)$ at the time $t=1.5$ [see Fig. 5(b)], $\Delta _{3,1x}^{\ast }$ is nearly in its
thermodynamic equilibrium [see Fig. 6(b)]. At the two
sides of the peak (valley), the system deviates from its equilibrium in
opposite directions. (ii) For each kind of TNE quantity, the shear
component, such as $\Delta _{2xy}^{\ast }$ and $\Delta _{4,2xy}^{\ast }$, or
the flux in the $x$ direction, such as $\Delta _{3,1x}^{\ast }$ and $\Delta
_{3xxx}^{\ast }$, own the largest amplitudes (other components such as $%
\Delta _{2xx}^{\ast }$, $\Delta _{3,1y}^{\ast }$, $\Delta _{3xxy}^{\ast }$,
etc., are not shown here). Behaviors of TNE measures can be interpreted as
follows. Physically, $\bm{\Delta}_{2}^{\ast }$ and $\bm{\Delta}_{3,1}^{\ast
} $ correspond to more generalized viscous stress and heat flux,
respectively; $\bm{\Delta}_{3}^{\ast }$ and $\bm{\Delta}_{4,2}^{\ast }$
correspond to more generalized fluxes of viscous stress and heat flux,
respectively. Essentially, KHI is a kind of shearing instability through which
mass-momentum-energy transfer through the initial interface,
%along the direction of the initial perturbation,
resulting in remarkable shear-induced nonequilibrium and transporting
nonequilibrium. (iii) For any TNE manifestation, although the nonequilibrium
amplitude evolves complicatedly, the nonequilibrium region (NER) extends
with evolution on account of the KHI-induced mixing process.
The width of the NER justly corresponds to the width of the mixing layers $w_{\text{mixing}}$.
Therefore, we present here an interface-tracking method through
tracking the leftmost and the rightmost positions with
$|\bm{\Delta}_{m,n}^{\ast }| \geq  |\bm{\Delta}_{m,n}^{\ast}|_{\text{th}}$, where $|\bm{\Delta}_{m,n}^{\ast}|_{\text{th}}$
is a threshold of $|\bm{\Delta}_{m,n}^{\ast }|$.

%%%%%%%%%%%%%%%%%%%%%%%%%%%%%%%%%%%%%%%%%%%%%%%%%%%%%%%%%%%%%%%%%%%%%%%%%%%%%%%%%
\begin{figure*}[b!]
	\centering
	\includegraphics[width=0.75\textwidth]{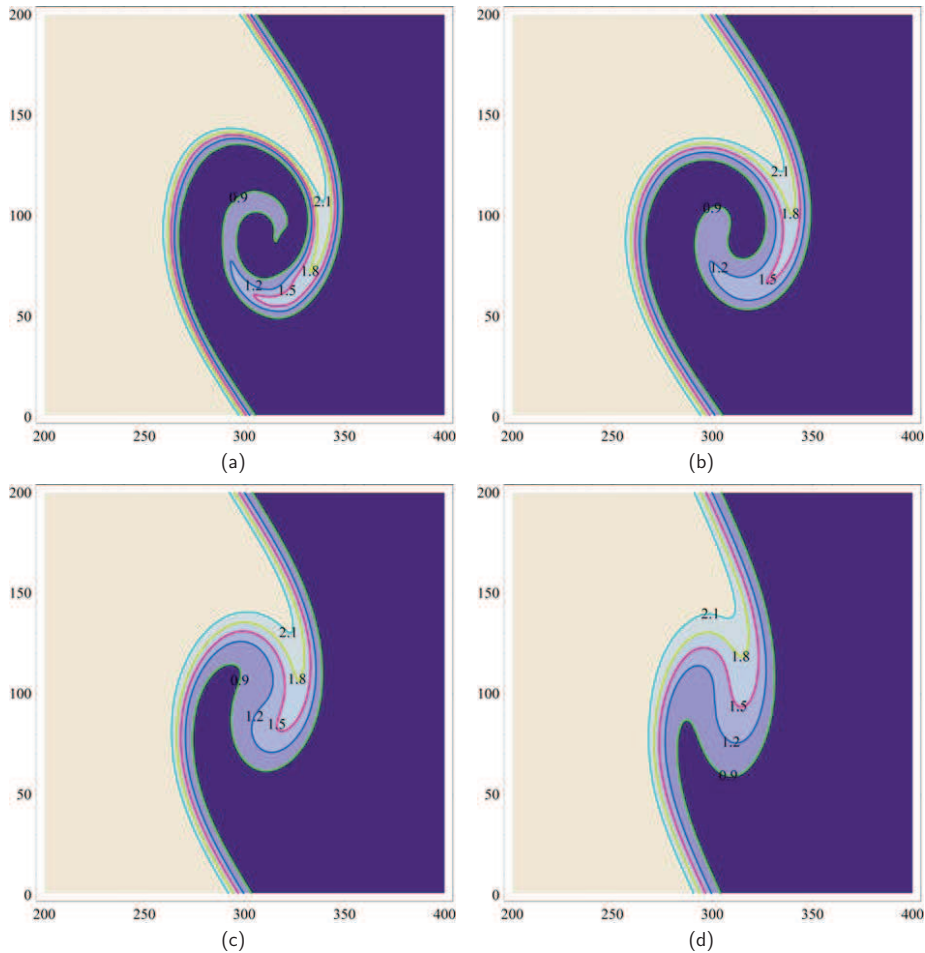}
	\caption{Vortices in the mixing layer as a function of
		viscosity at $t=1.4$, where $\tau=10^{-5}, 10^{-4}, 5\times 10^{-4}$
		and $10^{-3}$ in {\bf(a)}, {\bf(b)}, {\bf(c)} and {\bf(d)}, respectively. Only the part with $x$
		in the range $[200:400]$ of the full grid is shown. From purple to light
		brown, the density increases.}
\end{figure*}
%%%%%%%%%%%%%%%%%%%%%%%%%%%%%%%%%%%%%%%%%%%%%%%%%%%%%%%%%%%%%%%%%%%%%%%%%%%%%%%%%

%%%%%%%%%%%%%%%%%%%%%%%%%%%%%%%%%%%%%%%%%%%%%%%%%%%%%%%%%%%%%%%%%%%%%%%%%%%%%%%%%
\begin{figure*}[b!]
	\centering
	\includegraphics[width=0.92\textwidth]{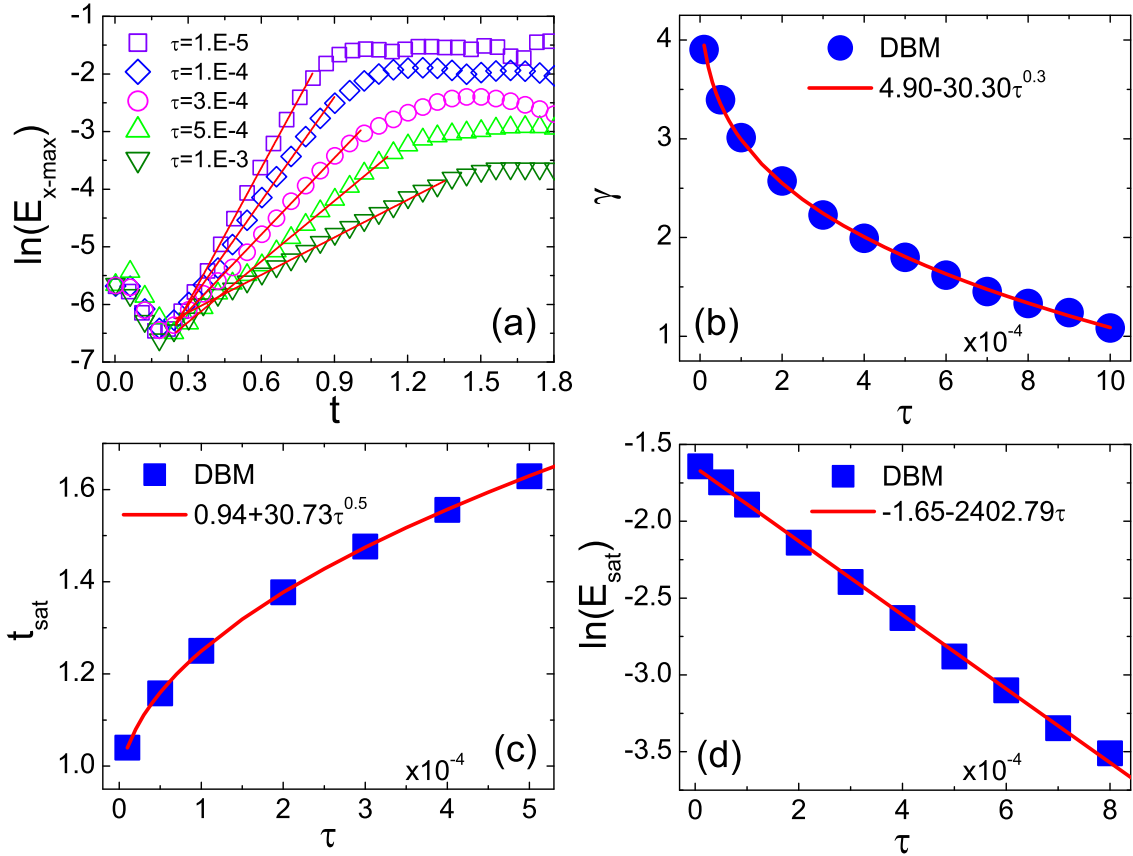}
	\caption{Time evolution of the logarithm of the perturbed
		peak kinetic energy $E_{x\text{-max}}$ for cases with various viscosities,
		where solid lines represent the linear fits to the initial linear growth
		regimes {\bf(a)}. Linear growth rate as a function of $\tau$ {\bf(b)}. Saturation time $t_{\text{sat}}$ {\bf(c)}
		and peak perturbed kinetic energy {\bf(d)} for cases with various viscosities.}
\end{figure*}
%%%%%%%%%%%%%%%%%%%%%%%%%%%%%%%%%%%%%%%%%%%%%%%%%%%%%%%%%%%%%%%%%%%%%%%%%%%%%%%%%

%%%%%%%%%%%%%%%%%%%%%%%%%%%%%%%%%%%%%%%%%%%%%%%%%%%%%%%%%%%%%%%%%%%%%%%%%%%%%%%%%
\begin{figure*}[b!]
	\centering
	\includegraphics[width=0.95\textwidth]{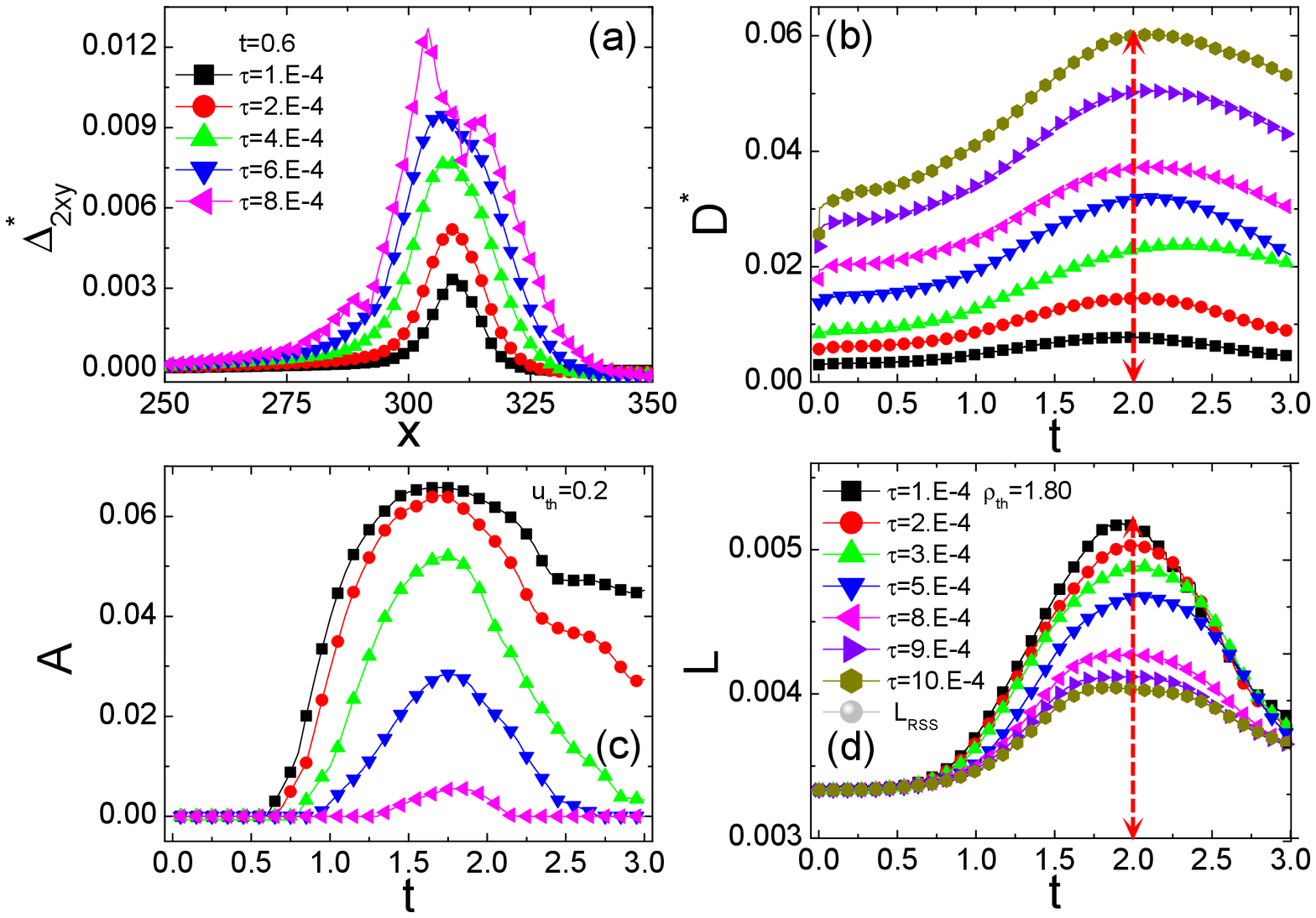}
	\caption{Effects of viscosity on nonequilibrium measure $
		\Delta_{2xy}^*$ {\bf(a)} and high velocity area fraction $A$ for $
		|u|_{\text{th}}=0.2$ {\bf(c)}. Effects of viscosity on the
		non-dimensionalized nonequilibrium intensity $D^*$ {\bf(b)} and boundary
		length $L$ for $\rho _{\text{th}}=1.80 $ {\bf(d)}.}
\end{figure*}
%%%%%%%%%%%%%%%%%%%%%%%%%%%%%%%%%%%%%%%%%%%%%%%%%%%%%%%%%%%%%%%%%%%%%%%%%%%%%%%%%

Essentially, the mixing zone width and its growth rate are of great significance in the study of the
hydrodynamic instability and turbulent mixing \cite{29,103,104,105,106,107,108}.
The time evolution of $w_{\text{mixing}}$, revealing
the mixing extent and efficiency,
is an important parameter to quantitatively
study the development of KHI.
Usually, for incompressible KHI, the measurement is
readily performed by tracing the constant density. Nevertheless, in the
compressible case, how to measure the mixing layer remains a thorny problem.
The interface-tracking approach based on TNE measures may shed some light on
identification, labelling and extraction of characteristic structures from
complex physical fields.
To elaborate on this point, Fig. 7 exhibits temporal evolution of the mixing
layers thickness by tracking boundaries of $\Delta_{3xxx}^*$ and $\Delta_{31x}^*$.
It is evident that, although acquired from different TNE measures, $w_{\text{mixing}}$
approximately overlap with each other, manifesting that TNE quantities have
been coupled with each other. In other words, gradient in one quantity, say
density gradient, can stimulate gradients in other quantities, say velocity
and temperature gradients. The non-monotonic $w_{\text{mixing}}$ can be approximately divided into four
stages, i.e., a drastic increase until $t=1.4$, a mild increase, a plateau,
and a steep decrease from about $t=2.5$ which are basically consistent with
scenarios in Figs. 4 and 5.
Moreover, it is found that, when $t<1.4$, $w_{\text{mixing}}$ dramatically increase with $t$
according the the following way
\begin{equation}
w_{\text{mixing}} = A t^2 + B \text{,}
\end{equation}
with $A=50$ and $B=3$, which is substantially different from the the Richtmyer-Meshkov instability \cite{105}.

\subsection{Effects of viscosity}

Here, focus is on how and to what extent viscosity affects the dynamic
patterns, nonequilibrium and morphological features. To this end, we have
performed comparative calculations with fixed thermal conductivity $%
\kappa_T=1.2\times 10^{-3}$, but varying viscosity coefficients $\mu$
through changing relaxation time $\tau$ over two decades.
Figure 8 shows density maps for $\tau=10^{-5}, 10^{-4}, 5\times 10^{-4}$ and
$10^{-3}$ with five contour lines at $t=1.4$, respectively. Apparently, the
striking differences in density patters demonstrate that the evolution of
KHI depends on viscosity strongly. For fixed initial conditions and model
parameters, %except for $\tau$ related to the kinematic viscosity,
the larger the viscosity, the weaker the KHI, and the later the vortex
appears.
The viscous force persistently extracts perturbation kinetic energy from the
fluids and erases progressively the small substructures. Therefore, we see
from the top row that, at lower viscosity, largely rolled billows consisting
of more windings, diverse length scales, %diverse range of length scales
and very sharp density interfaces, appear. Larger viscosity delays the
formation of the instability and inhibits the onset completely in cases with
extremely high values.
As observed in the bottom row of Fig. 8, cusps, instead of roll-up motions,
forms. Figure 8 indicates that the fluid viscosity tends to suppress and
limit the growth of the KHI.

The stabilizing effects of viscosity can be further confirmed by the
temporal evolution of logarithm of the perturbed peak kinetic energy $E_{x%
	\text{-max}}(t)$ for cases with various $\tau$
[see Fig. 9(a)]. For
each case, $E_{x\text{-max}}(t)$ experiences four stages as well, in
accordance with Figs. 4, 5 and 7. After an initial settling period, it increases
exponentially %that corresponds to the onset of the KHI
during the linear phase, until the reach of the saturation stage with the
saturation energy $E_{\text{sat}}$. $E_{\text{sat}}$, determined as the
first peak amplitude in $E_{x\text{-max}}(t)$ arrived at time $t_{\text{sat}}
$, can be used to measure the suppression level by viscosity and the
non-linear evolution of KHI. After that, we also see the finite-amplitude
oscillation in $E_{x\text{-max}}(t)$, owing to the generation and
development of the subharmonic modes. Subsequently, $E_{x\text{-max}}(t)$
decreases almost exponentially towards the initial perturbed level ($t>1.8$, not shown here).
%an initial settling period, an exponential growth stage, a saturation phase, continued by a highly nonlinear regime with almost exponentially decrease in $E_{x\text{-max}}(t)$.
% Taking the logarithm of $E_{x\text{-max}}$ of each time step,
The linear growth rate $\gamma$ can be obtained from the slope of the linear
function fitted to the growth phase, shown by the solid lines.
%Meanwhile, the saturation level $E_{max}$ is determined as the first maximum in $E_{x\text{-max}}$, which is reached at time $t_{text{sat}}$.
$E_{x\text{-max}}(t)$ represents the interacting strength of two different
fluids. At the same moment, the larger the viscosity, the lower the
perturbed peak kinetic energy, manifesting hindering effects of viscosity on
KHI. Figure 9 demonstrates that the viscosity effects are
threefold: significantly refrain both the linear growth rate $\gamma$ and
the saturation energy $E_{\text{sat}}$, prominently prolong the duration of
the linear growth stage $t_{\text{linear}}$ or postpone the saturation time $%
t_{\text{sat}}$, approximately in the following ways:
\begin{equation}
\gamma =a-b\tau ^{0.3}\text{,}
\end{equation}
\begin{equation}
\ln E_{\text{sat}}=c-d\tau \text{,}
\end{equation}
\begin{equation}
t_{\text{sat}}=e+f\tau ^{0.5}\text{,}
\end{equation}
with $a=4.90$, $b=30.30$, $c=-1.65$, $d=2402.79$, $e=0.94$, and $f=30.73$,
respectively. In the classical case where Euler equations dominate, the
linear growth rate reads $\gamma _{c}=k\sqrt{\rho _{1}\rho _{2}} \Delta v/(\rho _{1}+\rho _{2})$ \cite{47,48},
where $\Delta v$ is the shear velocity difference.
Higher viscous dissipation hampers the development of $\Delta v$ and removes
more kinetic energy gained from the shear velocity, resulting in a
smaller growth rate, a lower saturation energy, and a longer linear growth
stage.

Figure 10(a) reveals how viscosity affects the shear TNE component $\Delta_{2xy}^*$
along the line $y=N_y/2$ at $t=0.6$. To be seen is that,
viscosity enhances both the local and global nonequilibrium strengths,
broadens the nonequilibrium region. For case with $\tau=10^{-4}$, the
nonequilibrium region is confined to $[294:322]$ and the nonequilibrium
amplitude is about $0.003$; while for case with $\tau=8\times 10^{-4}$, the
counterparts are $[260:342]$ and $0.0127$, respectively. Moreover, we find
that, although $\Delta_{2xy}^*$ is proportional to $\tau$, the amplitudes of
$\Delta_{2xy}^*$ do not increase with $\tau$ linearly. This is because, the
relaxation time $\tau$ plays opposite roles in both the thermodynamical and
hydrodynamical aspects. %(on...sides)
Thermodynamically, it maintains the system in a far-from-equilibrium state
through retaining the macroscopic quantity gradients to high levels. But
hydrodynamically, it extends the density profile, reduces the temperature
gradient, etc., then makes the system deviate less from its thermodynamic
equilibrium.
Effects of viscosity on high-velocity area fraction $A$ for $u_{\text{th}%
}=0.2$ are shown in Fig. 10(c). These curves behave qualitatively
similar and can be divided into three stages, corresponding to the time
delay stage with nearly zero value of $A$, the rapid increase and the
subsequent decrease in $A$. Asides from similarity, the distinct differences
resulted from various $\tau$ are as below. The larger the viscosity, the
longer the time delay for observable KHI under fixed velocity threshold,
%The larger the viscosity,
as well as the smaller the slope and amplitude of $A(t)$ curve in the second
stage. In fact, slopes of $A(t)$ curves correspond approximately to the
evolution speed of KHI. From this point of view, the instability is
remarkably decelerated by viscosity.

Figure 10(b) exhibits effects of viscosity on the
non-dimensionalized nonequilibrium intensity $D$ and boundary
length $L$ for $\rho _{\text{th}}=1.80$ [panel (d)]. Here $%
D$ is defined as
\begin{equation}
D=\frac{1}{L_{x}L_{y}}\int_{0}^{L_{x}}\int_{0}^{L_{y}}(\frac{\bm{\Delta}
	_{2}^{\ast 2}}{T^{2}}+\frac{\bm{\Delta }_{3,1}^{\ast 2}}{T^{3}}+\frac{
	\bm{\Delta}_{3}^{\ast 2}}{T^{3}}+\frac{\bm{\Delta }_{4,2}^{\ast 2}}{
	T^{4}})^{1/2}dxdy\text{,}
\end{equation}%
where $D=0$ in the thermodynamic equilibrium state and $D>0$ in the
thermodynamic nonequilibrium state. Clearly, for a fixed $\tau $, $D(t)$
increases to its maximum then decreases slowly. For cases with different $%
\tau $, all $D(t)$ curves increase with $\tau $ and approach their maxima at
about the same time $t=2.0$. The evolutions of $D(t)$ and $L(t)$ present
considerably high degree of correlation. For instance, $L(t)$ also increases
with $\tau $ and behaves the first increase and latter decrease way as $D(t)$%
. Once again, all $L(t)$ curves reach their peaks also at the same moment
with $D(t)$ for cases with various $\tau $. Physically, the dominating part
of non-dimensionalized nonequilibrium intensity $D(t)$ is a combination of
products of hydrodynamic quantities gradients ($\bm{\nabla}\rho $, $%
\bm{\nabla}\mathbf{u}$ and $\bm{\nabla}T$) and $\tau $. Therefore, for a
single curve, a larger boundary length $L(t)$ corresponds to greater density
gradients and more intense nonequilibrium extent. Owing to the violent
resistance effects of $\tau $ on KHI, we mention that, the smaller the $\tau
$, the larger the $L(t)$. Oppositely, the larger the $L(t)$, the weaker the $%
D(t)$, indicating the leading role of $\tau $ versus ($\bm{\nabla}\rho $, $%
\bm{\nabla}\mathbf{u}$ and $\bm{\nabla}T$). Comparisons between panels (b) and (d) show that the evolutions of TNE intensity and boundary length present high correlation and attain their maxima almost simultaneously.

%%%%%%%%%%%%%%%%%%%%%%%%%%%%%%%%%%%%%%%%%%%%%%%%%%%%%%%%%%%%%%%%%%%%%%%%%%%%%%%%%
\begin{figure*}[b!]
	\centering
	\includegraphics[width=0.6\textwidth]{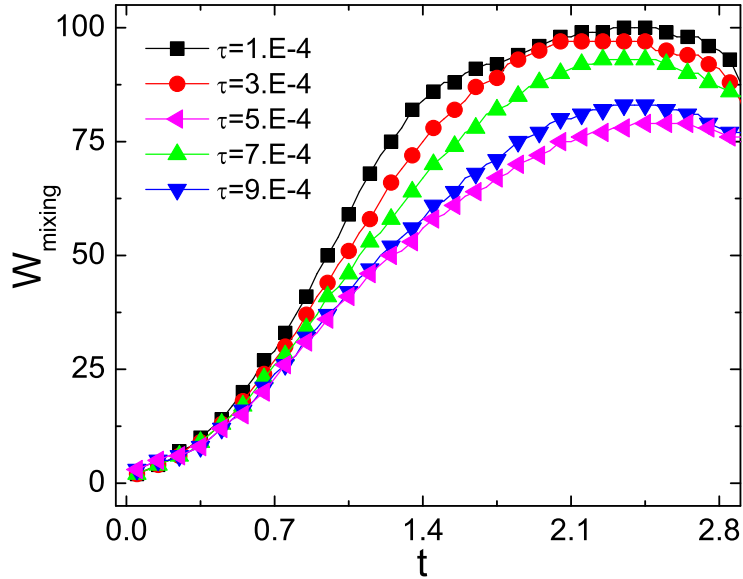}
	\caption{Widths of the mixing layers for cases with various viscosities obtained from evolutions of $\Delta_{3xxx}^*$.}
\end{figure*}
%%%%%%%%%%%%%%%%%%%%%%%%%%%%%%%%%%%%%%%%%%%%%%%%%%%%%%%%%%%%%%%%%%%%%%%%%%%%%%%%%

Figure 11 displays widths of the mixing layers $w_{\text{mixing}}$ for several cases with different viscosities obtained from the interface-tracking technique based on TNE measure $\Delta_{3xxx}^*$.
As shown, for different cases, evolutions of $w_{\text{mixing}}$ are self-similar and
behave qualitatively quite similar as boundary length $L$ [Fig. 10(d)].
Nevertheless, different relaxation times $\tau$ generate different $w_{\text{mixing}}$.
The larger the $\tau$, the narrower the $w_{\text{mixing}}$. Figure 11 quantitatively reflects the hindering effects of viscosity on the
onset and development of KHI.

\subsection{Effects of heat conduction}

%%%%%%%%%%%%%%%%%%%%%%%%%%%%%%%%%%%%%%%%%%%%%%%%%%%%%%%%%%%%%%%%%%%%%%%%%%%%%%%%%
\begin{figure*}[b!]
	\centering
	\includegraphics[width=0.75\textwidth]{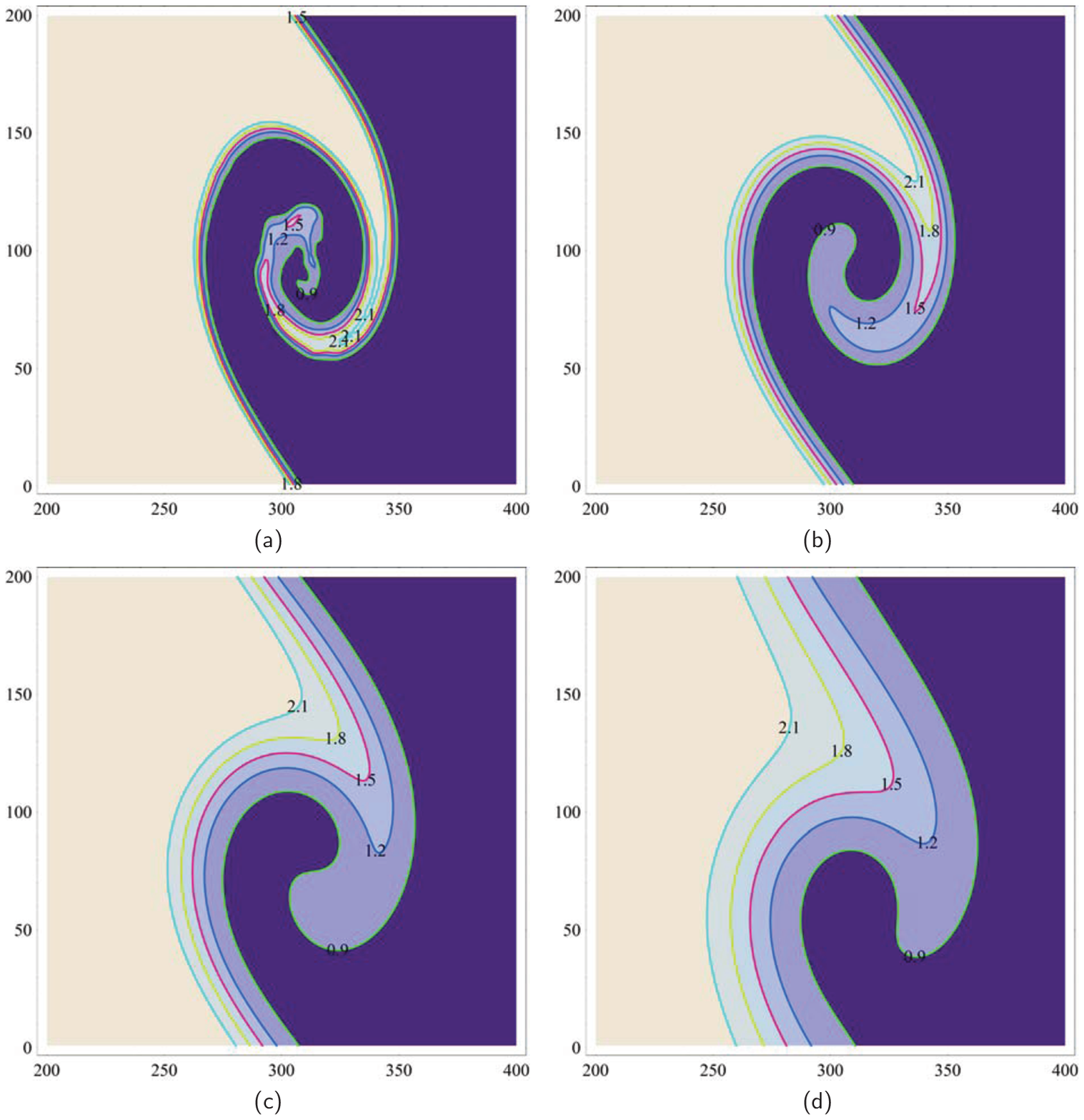}
	\caption{ Vortices in the mixing layer as a function of heat
		conduction at $t=1.4$, where $\kappa_T=1.5\times 10^{-4}, 1.5\times
		10^{-3}, 6\times 10^{-3}$ and $1.5\times 10^{-2}$ in {\bf(a)}, {\bf(b)}, {\bf(c)} and {\bf(d)}, respectively. Only the part with $x$ in the range $[200:400]$ of the full
		grid is shown. From purple to light brown, the density increases.}
\end{figure*}
%%%%%%%%%%%%%%%%%%%%%%%%%%%%%%%%%%%%%%%%%%%%%%%%%%%%%%%%%%%%%%%%%%%%%%%%%%%%%%%%%

Effects of heat conduction are tentatively analyzed in a similar way with
fixed relaxation time $\tau=10^{-4}$ and various thermal conductivities $%
\kappa_T$.
%The initial conditions and other parameters are in accordance with that used in Fig. 4.
Shown in Fig. 12 are the typical density patterns at $t=1.4$ for cases with $%
\kappa_T=1.5\times 10^{-4}, 1.5\times 10^{-3}, 6\times 10^{-3}$ and $%
1.5\times 10^{-2}$, respectively. The KH structure changes considerably as
the heat conduction varies. For case with smaller $\kappa_T$, the cat's eye
structure with spiral arm has been observed from the density fields. While, at
larger $\kappa_T$, the initial disturbance merely develops into brawny cusp
that tentatively whirls across the transition layer. Meanwhile, the mixing
layer is thickening and the higher order harmonic is refraining by thermal
diffusion, which results in single and larger scale configuration. Roughly
speaking, the heat conduction stabilize the KHI by suppressing the growth of
the initial perturbation and the appearance of the higher order harmonics.

%%%%%%%%%%%%%%%%%%%%%%%%%%%%%%%%%%%%%%%%%%%%%%%%%%%%%%%%%%%%%%%%%%%%%%%%%%%%%%%%%
\begin{figure*}[b!]
	\centering
	\includegraphics[width=0.6\textwidth]{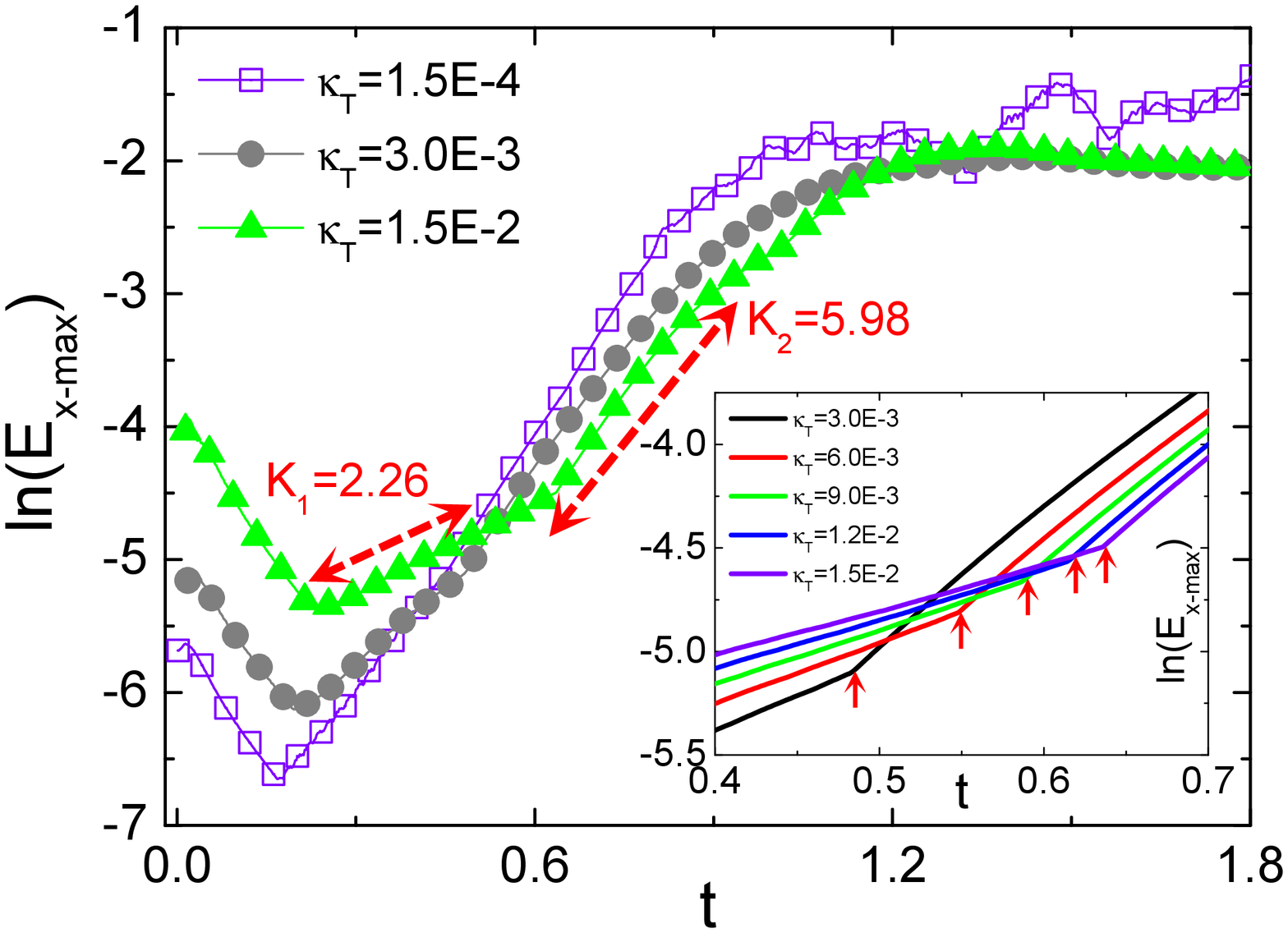}
	\caption{ Time evolution of the perturbed peak kinetic
		energies $E_{x\text{-max}}$ in a ln-linear scale for cases with various heat
		conductivities $\kappa_T$. Evolution of $E_{x\text{-max}}(t)$ during
		the stage $0.4<t<0.7$ has been shown in the inset, where red arrows pointing
		towards the inflection points.}
\end{figure*}
%%%%%%%%%%%%%%%%%%%%%%%%%%%%%%%%%%%%%%%%%%%%%%%%%%%%%%%%%%%%%%%%%%%%%%%%%%%%%%%%%

%%%%%%%%%%%%%%%%%%%%%%%%%%%%%%%%%%%%%%%%%%%%%%%%%%%%%%%%%%%%%%%%%%%%%%%%%%%%%%%%%
\begin{figure*}[b!]
	\centering
	\includegraphics[width=0.82\textwidth]{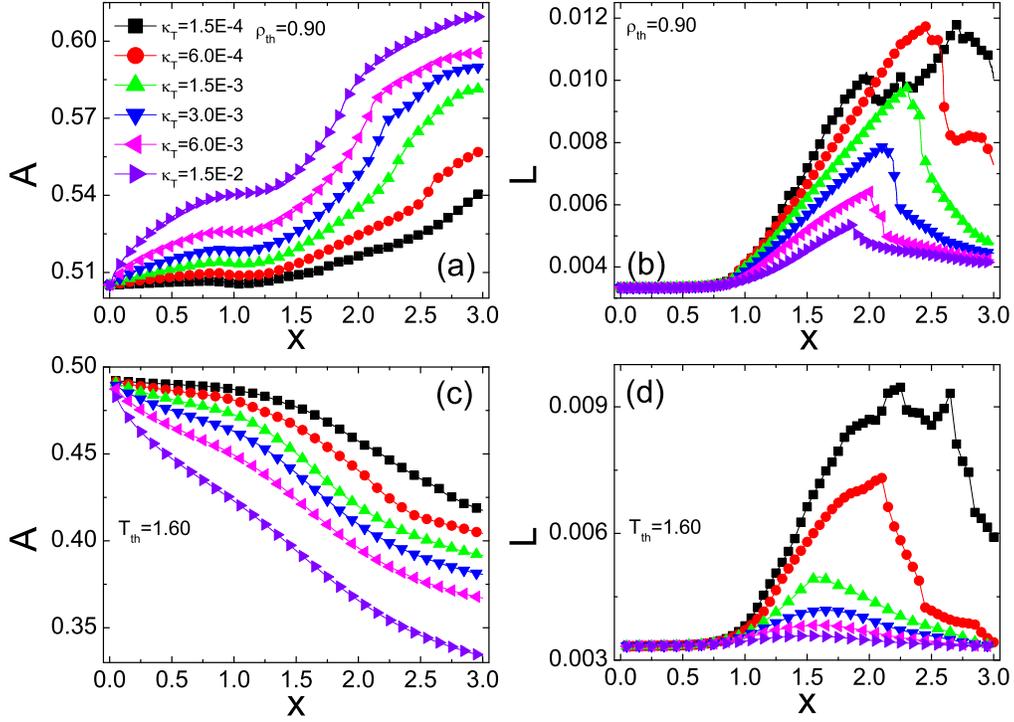}
	\caption{ Effects of heat conduction on high density area
		fraction $A$ {\bf(a)} and high density boundary length $L$ {\bf(b)} for
		$\rho_{\text{th}}=0.90$. Effects of heat conduction on high
		temperature area fraction $A$ {\bf(c)} and high temperature boundary
		length $L$ {\bf(d)} for $T_{\text{th}}=1.60$. }
\end{figure*}
%%%%%%%%%%%%%%%%%%%%%%%%%%%%%%%%%%%%%%%%%%%%%%%%%%%%%%%%%%%%%%%%%%%%%%%%%%%%%%%%%

%%%%%%%%%%%%%%%%%%%%%%%%%%%%%%%%%%%%%%%%%%%%%%%%%%%%%%%%%%%%%%%%%%%%%%%%%%%%%%%%%
\begin{figure*}[b!]
	\centering
	\includegraphics[width=0.98\textwidth]{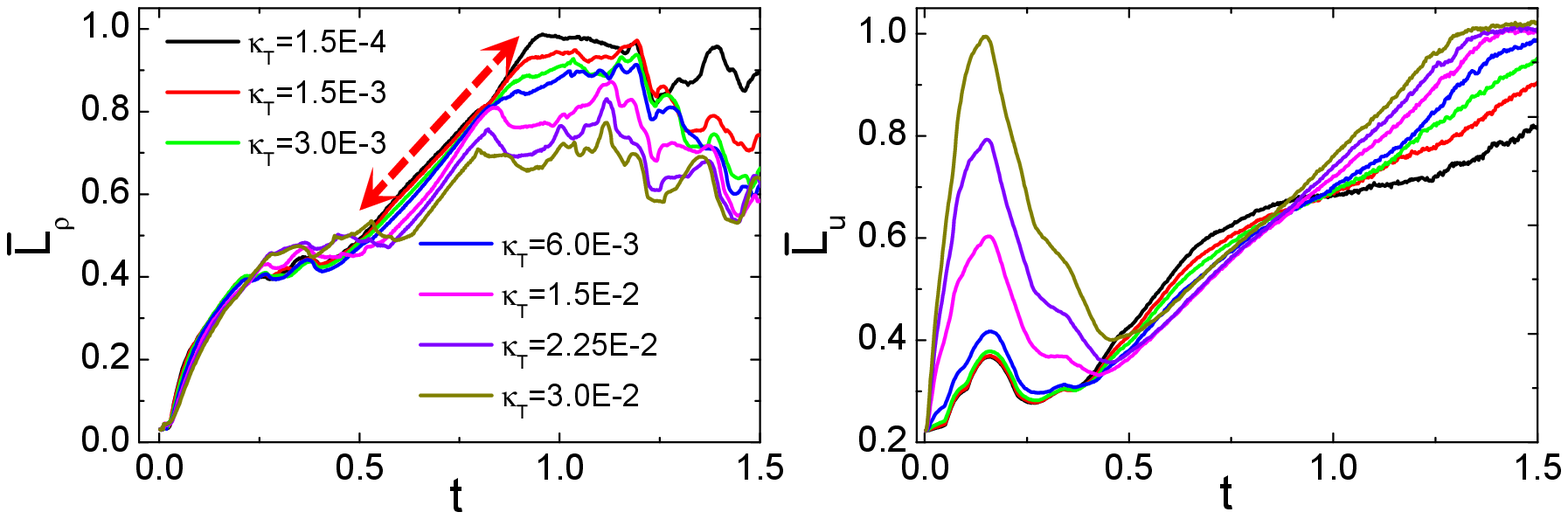}
	\caption{ Temporal evolutions of the average widths of
		density {\bf(a)} and longitudinal velocity {\bf(b)} transition layers. }
\end{figure*}
%%%%%%%%%%%%%%%%%%%%%%%%%%%%%%%%%%%%%%%%%%%%%%%%%%%%%%%%%%%%%%%%%%%%%%%%%%%%%%%%

Figure 13 gives time evolution of the perturbed peak kinetic energies $E_{x%
	\text{-max}}$ in a ln-linear scale for cases with various $\kappa_T$.
Generally speaking, $\kappa_T$ reduces the linear growth rate. Nevertheless,
the stabilizing effects are not as obvious as viscosity, moreover, become
less important for larger $\kappa_T$, manifested by %reflected in
the little differences in slopes of $E_{x\text{-max}}(t)$ and the nearly
identical saturation kinetic energy $E_{\text{sat}}$ for diverse cases.
Meanwhile we mention a significant effect of $\kappa_T$, for large enough $%
\kappa_T$, say $\kappa_T=1.5\times10^{-2}$, after the initial transient
period and before the vortex has been well formed, the evolution of $E_{x%
	\text{-max}}(t)$ undergoes two linear stages with distinct slopes. To
highlight this point, evolution of $E_{x\text{-max}}(t)$ during the stage $%
0.4<t<0.7$ has been shown in the inset, where red arrows pointing towards
the inflection points. The larger the heat conductivity, the later the
inflection point appears, the smaller the perturbed peak kinetic energy
before the inflection point, but the greater the perturbed peak kinetic
energy after the inflection point. Behaviors of $E_{x\text{-max}}(t)$ at
large $\kappa_T$ demonstrate that the heat conduction effects are not
monotonic: first refrain but favor the evolution afterwards.

To further clarify effects of heat conduction, we illustrate the
morphological features of density and temperature fields in Fig. 14. For all
cases, the high density area fraction $A(t)$ and boundary length $L(t)$ for $%
\rho _{\text{th}}=0.90$ increase with time due to the viscous heating, the
timely heat diffusion, and the appearance of new interface between fluids of
different densities. Nevertheless, effects of heat conduction on $A(t)$ and $%
L(t)$ are completely opposite. At the same moment, the larger the $\kappa
_{T}$, the higher the $A $, but the lower the $L$. This indicates that the
thermal diffusion primarily %(wholly)
promotes the translational motion of the interface and suppresses the curl
of the initial perturbation. The slope of $L(t)$ clearly represents the
development rate of KHI. Once again, we conclude the KHI is decelerated by
heat conduction. The $A(t)$ and $L(t)$ curves for temperature field at
threshold $T_{\text{th}}=1.60$ are shown in the bottom row of Fig. 14.
Similar conclusion can be acquired from $L(t)$ curves for temperature field.
But for $A(t)$ curves, we see that heat conduction lowers the proportion
with temperature $T>T_{\text{th}}$ owing to the larger $\kappa _{T}$ that
makes the system approach thermodynamical equilibrium quickly.

To interpret behaviors of $E_{x\text{-max}}(t)$ at large $\kappa_T$,
in Fig. 15, we monitor the temporal histories of the normalized widths of density and velocity transition layers, refereed to as $\bar{L}_{\rho}$ and $\bar{L}_{u}$, respectively. Therein, $\bar{L}_{\rho}$ is defined as
\begin{equation}
\bar{L}_{\rho} =  \frac{1}{L_{\rho}^{\text{max}}} \frac{1}{L_{x} L_{y} }\int_{0}^{L_{x}}\int_{0}^{L_{y}}\rho (
|\frac{\partial \rho}{\partial x}|^{-2}
+|\frac{\partial \rho }{\partial y}|^{-2})^{1/2}dxdy
\end{equation}
when $|\frac{\partial \rho }{\partial x}|>0.1$ and $|\frac{\partial \rho }{%
	\partial y}|>0.1$;
otherwise $\bar{L}_{\rho}=0$. $\bar{L}_{u}$ is defined similarly.
As can be seen, for all $\bar{L}_{\rho}(t)$ curves, when $t<0.25$, they increase sharply and overlap with each other, representing the initial relaxation stage where
the violent temperature gradients attempt to initiate the
instability through thickening the density transition layer.
% without conspicuous instability for all cases.
Compared to $\bar{L}_{\rho}(t)$, $\bar{L}_{u}(t)$ grows more steeply, impetuously and inconsistently during this period. As a result, significant differences appear in $\bar{L}_u(t)$.
For example, $\bar{L}_u(t)$ for $\kappa_T=3\times 10^{-2}$ is about 3 times wider than that for $\kappa_T=1.5\times 10^{-4}$ at $t=0.14$.
%the larger the $\kappa_T$, the wider the $\bar{L}_{\rho}$.
The wider transition layer $\bar{L}_{u}$ effectively decreases the local shear velocity difference $\Delta v$ and results in a smaller linear growth rate in $E_{x-\text{max}}(t)$.
This can be seen more clearly from the square of the linear growth rate for the classical case \cite{48}, $\gamma_{c}^{2}=k^{2}\rho _{1}\rho_{2}(v_{1}-v_{2})^{2}/(\rho_{1}+\rho_{2})^{2}\propto (1-A^{2})\Delta v^{2}$, where $A=(\rho_{1}-\rho_{2})/(\rho_{1}+\rho _{2})$ is the Atwood number.

On the other hand, when $0.25<t<1.0$, all $\bar{L}_{\rho}(t)$ curves further increase with time, but decreases prominently with $\kappa_T$, until the arrivals of the slightly oscillating platforms with various heights (labelled by red line with double-headed arrows).
The persistent and notable growth in $\bar{L}_{\rho}$,
gives rise to a wider density transition zone that reduces the effective Atwood number around the interface.
Therefore, in the process of exchanging momentum along the direction normal to the interface, the perturbation obtains energy easily from the shear layers than in cases with steeper interfaces or with higher density ratios.
% and results in a larger growth rate.
Meanwhile, $\bar{L}_{u}$ slide to their valleys at about $t=0.5$.
Consequently, the limited $\bar{L}_u$ and the fully developed $\bar{L}_{\rho}$ advance the instability enormously, and give
rise to the second and steeper increase in $E_{x-\text{max}}(t)$.

\section{Conclusions and remarks}

An efficient and easily implementable discrete Boltzmann model is proposed to systematically study the viscosity and heat conduction effects on the onset and growth of compressible Kelvin-Hemholtz instability (KHI).
Both thermodynamic nonequilibrium (TNE) and morphological characterizations are extracted, for the first time,  to analyze and understand the configurations and the kinetic processes.

The findings are as below.
On the technical side, two novel approaches, independent from each other, are presented to quantitatively feature the evolution of KHI.
One is to determine the thickness of mixing layers via tracking the distributions and evolutions of the TNE measures.
The other is to access the growth rate of KHI via the slopes of Minkowski measures.
On the physical side, it is interesting to find that the time histories of width of mixing layer, TNE intensity, and boundary length between the high and low macroscopic quantity regimes show high correlation and attain their maxima simultaneously.
The effects of viscosity are twofold. One is to stabilize the KHI through reducing the linear growth rate, prolonging the duration of the linear growth stage, and suppressing the hydrodynamic velocity and the perturbed peak kinetic energy.
The other is to enhance both the local and global nonequilibrium strengths via enlarging the relaxation time and broadening the nonequilibrium region,
respectively.
Contrary to the monotonically inhibiting effects of viscosity, the simulations reveal that the heat conduction effects refrain at first then accelerate the evolution afterwards.
This is because heat conduction extends
both the widths of density and velocity transition layers simultaneously. While the two kinds of widths act oppositely on the evolution of KHI. During the first period, the growing velocity transition layer dominates the evolution; After that,
the persistently increasing density transition layer together with
the temporarily decreasing velocity transition layer dominates the evolution jointly.

Although this work focus on two-dimensional case, the introduced morphological analysis method and the  developed interface-capturing technique based on TNE measures, can be applied to pick up information from  three-dimensional physical fields. The nonequilibrium and morphological characterizations of three-dimensional KHI in multiphase flows and compressible flows deviating far away from thermodynamic equilibrium deserve further study and are currently in progress.

\section*{Acknowledgments}

YG, CL, HL and ZL acknowledge the support from the National Natural Science Foundation of China
(Grants No. 11875001, No. 51806116, and No. 11602162),
Natural Science Foundation of Hebei Province (Grants No. A2017409014 and No.
2018J01654), Natural Science Foundations of Hebei Educational Commission (Grant No. ZD2017001).
AX and GZ acknowledge the support from the National Natural Science Foundation of China
(Grant No. 11772064), CAEP Foundation (Grant No. CX2019033),
Science Challenge Project (Grant No. JCKY2016212A501),
the opening project of State Key Laboratory of Explosion Science and Technology (Beijing Institute of Technology, Grant No. KFJJ19-01M).

\end{document}